\documentclass[11pt]{article}
\pdfoutput=1

\usepackage{a4wide}
\usepackage[fleqn]{amsmath}

\RequirePackage{amsmath,amssymb,amsxtra,amsthm}

\RequirePackage[T1]{fontenc}
\RequirePackage[utf8]{inputenc}

\RequirePackage{mathrsfs}
\RequirePackage{mathpazo}

\RequirePackage{setspace}
\RequirePackage{array}
\RequirePackage{booktabs}

\RequirePackage{braket}

\RequirePackage[pdftex,final]{graphicx}
\graphicspath{{Plots/}}

\newcommand{\be}{\begin{equation}}
\newcommand{\ee}{\end{equation}}
\newcommand{\bea}{\begin{eqnarray}}
\newcommand{\eea}{\end{eqnarray}}
\newcommand{\IR}{\mathbb{R}}

\newcommand{\IZ}{\mathbb{Z}}

\newcommand{\cF}{\mathcal{F}}

\newcommand{\cA}{\mathcal{A}}

\newcommand{\cS}{\mathcal{S}}
\newcommand{\cI}{\mathcal{I}}

\newcommand{\cM}{\mathcal{M}}
\newcommand{\cW}{\mathcal{W}}
\newcommand{\cN}{\mathcal{N}}
\newcommand{\cE}{\mathcal{E}}

\newcommand{\cZ}{\mathcal{Z}}
\newcommand{\cO}{\mathcal{O}}
\newcommand{\cR}{\mathcal{R}}
\newcommand{\cT}{\mathcal{T}}

\newcommand{\de}{\mathrm{d}}

\newcommand{\I}{\mathrm{i}}

\newcommand{\ar}[2]{{\textstyle \big[{#1\atop #2}\big]}}
\newcommand{\sgn}{{\rm sgn}}

\renewcommand{\mod}{\,{\rm mod}\,}

\numberwithin{equation}{section}
\numberwithin{table}{section}
\numberwithin{figure}{section}

\author{
  \begin{minipage}{0.97\linewidth}
    \vspace{1cm}
    \begin{center}
      \begin{small}
        \textbf{Carlo Angelantonj}$^{1}$, \textbf{Ioannis Florakis}$^{2}$ and  \textbf{Boris Pioline}$^{3,4}$
     \end{small}
    \end{center}
    \vspace{.3cm} \hspace{1.3cm}\begin{minipage}{.75\linewidth}
      {\it \begin{footnotesize}
          \begin{itemize}
          \item[${}^1$] Dipartimento di Fisica, Universit\`a di Torino, and INFN Sezione di Torino
          \\
            Via P. Giuria 1, 10125 Torino, Italy
          \item[${}^2$] Max-Planck-Institut f\"{u}r Physik,\\
	    Werner-Heisenberg-Institut,
          80805 M\"{u}nchen, Germany
          \item[${}^3$] CERN Dep PH-TH, 1211 Geneva 23, Switzerland
         \item[${}^4$] Laboratoire de Physique Th\'eorique et Hautes Energies, CNRS UMR 7589,
         \\
         Universit\'e Pierre et Marie Curie - Paris 6, 4 place Jussieu,
         75252 Paris cedex 05, France
          \end{itemize}
        \end{footnotesize}}
    \end{minipage}
    \vspace{1cm}
  \end{minipage}
}

\date{}

\title{\vspace{3cm}
  \begin{huge}
    \textbf{Rankin-Selberg methods for \\[2mm]  closed strings on orbifolds}
  \end{huge}
}

\begin{document}

\begin{titlepage}
  \maketitle
  \thispagestyle{empty}

  \vspace{-14cm}
  \begin{flushright}
    CERN-PH-TH/2013-049\\  
    MPP-2013-102
   \end{flushright}

  \vspace{11cm}

  \begin{center}
    \textsc{Abstract}\\
  \end{center}

In recent work we have developed a new unfolding method for computing one-loop modular integrals in string theory involving the Narain partition function and, possibly, a weak almost holomorphic elliptic genus. Unlike the traditional approach, the Narain lattice does not play any role in the unfolding procedure,  T-duality is kept manifest at all steps, a choice of Weyl chamber is not required and the analytic structure of the amplitude is transparent. In the present paper, we generalise this procedure to the case of Abelian $\IZ_N$ orbifolds, where the integrand decomposes into a sum of orbifold blocks that can be organised into orbits of the Hecke congruence subgroup $\Gamma_0(N)$. As a result, the original modular integral reduces to an integral over the fundamental domain of $\Gamma_0(N)$, which we then evaluate by extending our previous techniques. Our method is applicable, for instance, to the evaluation of one-loop corrections to BPS-saturated couplings in the low energy effective action of closed string models, of quantum corrections to the K\"ahler metric and, in principle, of the free-energy of superstring vacua.

\vfill
{\small
\begin{itemize}
\item[E-mail:] {\tt carlo.angelantonj@unito.it}\\ {\tt florakis@mppmu.mpg.de}\\
{\tt boris.pioline@cern.ch}
\end{itemize}
}
\vfill

\end{titlepage}

\setstretch{1.1}

\tableofcontents


\section{Introduction}

Perturbative one-loop computations in closed string theory typically require integrating over the moduli space of conformal structures on the world-sheet torus. Since the latter is isomorphic to the Poincar\'e upper half-plane $\mathbb{H}$ quotiented by the full modular group $\Gamma={\rm SL}(2;\IZ)$, the one-loop amplitude is  expressed as an integral $\int_{\cF} \de\mu \, \cA(\tau)$ of a modular invariant function over the fundamental domain $\cF =\Gamma\backslash \mathbb{H}$. The evaluation of such integrals is a mandatory step in any attempt to extract quantitative predictions from closed string models.

The standard approach to evaluating such integrals, known as the orbit method, relies on representing the integrand function as a (combination of) Poincar\'e series, {\em i.e.} a sum over ${\rm SL} (2;\mathbb{Z})$ images. It is important that such series be absolutely convergent, since only in this case can one trade the infinite sum over images for an integration over a semi-infinite strip, thus converting a stringy one-loop integral into a  conventional  field-theoretical Schwinger integral.
The technical task of identifying a suitable Poincar\'e series is not in general straightforward and  depends on the properties of the automorphic function $\cA (\tau)$. 

In string theory compactifications, the automorphic function $\cA(\tau )$ often decomposes into the product $\varGamma_{(d,d+k)}\, \varPhi$ of 
 the partition function $\varGamma_{(d,d+k)}$ of a Narain lattice with signature $(d,d+k)$, that encodes the dependence of the physical couplings on the compactification moduli, times a model-dependent  modular form $\varPhi$ of weight $-k/2$. In these cases, after Poisson resummation on the momenta,
 one may cast the Narain lattice partition function into a sum of  Poincar\'e series of ${\rm SL} (2;\mathbb{Z})$ suited for the unfolding of $\cF$. Although this approach has been applied with success in the past \cite{O'Brien:1987pn, McClain:1986id,Dixon:1990pc}, it suffers from certain drawbacks:  the resulting Poincar\'e series representation is absolutely convergent throughout the integration domain only in restricted regions of the Narain moduli space,   the decomposition  into ${\rm SL} (2;\mathbb{Z})$ orbits spoils the manifest T-duality invariance, and  the analytic properties of the physical couplings are obscured.

Recently, we proposed a new approach to evaluating such one-loop integrals that overcomes the aforementioned problems \cite{Angelantonj:2011br,Angelantonj:2012gw}. 
The main idea is to represent the function $\varPhi$ in terms of an absolutely convergent Poincar\'e series and use \emph{it}, rather than the lattice partition function, for the unfolding procedure. In practice, as shown in \cite{Angelantonj:2011br,Angelantonj:2012gw}, for the cases where $\varPhi$ is a weak almost holomorphic modular form, one may always obtain such a representation in terms of 
absolutely convergent Niebur-Poincar\'e series \cite{0288.10010,0543.10020}. Alternatively, in the case where $\varPhi$ is of moderate growth at $\infty$, one can resort to the Rankin-Selberg method \cite{0021.39201,0023.22201,MR656029} commonly used in Mathematics  to extract the analytic properties of $L$-series. The advantage of this approach is that it yields a manifestly T-duality invariant representation of the one-loop amplitude, which is valid throughout the Narain moduli space and whose singularity structure is transparent. While the occurrence of the Narain partition function is typical in compactifications on $d$-dimensional tori, the (almost) holomorphy of $\varPhi$ usually holds  for  amplitudes protected by supersymmetry, such as threshold corrections to gauge and gravitational couplings in certain superstring vacua \cite{Dixon:1990pc,Antoniadis:1992rq,Harvey:1995fq,Harvey:1996gc}. 

In many cases however, including closed-string vacua on orbifolds, 
the integrand $\cA(\tau)$ naturally decomposes into a finite sum of contributions, associated to the various orbifold sectors, which are related by  elements in ${\rm SL} (2;\mathbb{Z})$. Schematically, one  faces  one-loop integrals of the form
\begin{equation}
\int_{\cF} \de\mu\, \cA_{\rm orbifold} (\tau ) = \int_\cF \de\mu\, \sum_{\gamma \in \Gamma_N \backslash \Gamma } \cA_N (\gamma \tau )\,,
\label{introcoset}
\end{equation}
where $\Gamma_N \subset \Gamma$ is a level-$N$ congruence subgroup, and $\cA_N$ is an automorphic function under $\Gamma_N$, represented, in most cases of interest, by the product of a (possibly shifted) Narain partition function times some modular form $\varPhi_N$ of suitable weight with respect to $\Gamma_N$. In these cases, it is advantageous to first unfold $\cF$ into the fundamental domain $\cF_N$ of $\Gamma_N$ by using the coset decomposition \eqref{introcoset}, leading to the integral
\be
\int_{\cF_N} \de\mu \, \cA_N (\tau )\,,
\label{introcosetN}
\ee
and then apply the orbit method to unfold $\cF_N$. This approach was first discussed in the Physics literature in \cite{Mayr:1993mq} where the non-universality of gauge threshold corrections for heterotic orbifold compactifications with non-factorisable tori was shown, and was later applied to the computation of quartic gauge couplings \cite{Lerche:1998nx,Lerche:1998gz}, to freely-acting orbifolds with partial or total supersymmetry breaking in \cite{Angelantonj:2006ut,Kiritsis:1997ca, trapletti}, as well as in the context of $\mathcal{N}=4$ topological amplitudes \cite{Hohenegger:2011ff}.  These papers rely on unfolding the fundamental domain $\cF_N$  against the Narain lattice using the traditional implementation of the orbit method, and thus suffer from the same drawbacks outlined above. As we proposed in \cite{Angelantonj:2011br,Angelantonj:2012gw}, these drawbacks can  be circumvented by representing $\varPhi_N$ in terms of an absolutely convergent Poincar\'e series of $\Gamma_N$, and using it to unfold $\cF_N$. To this end, in this paper we generalise the construction of  Niebur-Poincar\'e series to congruence subgroups and use them to represent any weak almost holomorphic modular form of $\Gamma_N$. This allows one to evaluate the integral \eqref{introcosetN} at any point in the Narain moduli space, while keeping T-duality manifest. Similarly, we shall generalise the Rankin-Selberg method to the case of congruence subgroups, in order to treat functions of moderate growth\footnote{See \cite{cardella} for an earlier attempt at applying the Rankin-Selberg method to strings on orbifolds.}. 

In applications to Abelian $\mathbb{Z}_N$ closed-string orbifolds, the coset decomposition \eqref{introcoset} involves the congruence subgroup $\Gamma_1 (N)$, defined in Section \ref{sec_orbifold}. For $N$ prime,  the index 
of $\Gamma_1(N)$ inside the full modular group is $N^2-1$, which agrees with the number of non-trivial orbifold sectors. When $N$ is not prime, the decomposition \eqref{introcoset} involves instead several independent orbits, each one associated to a congruence subgroup $\Gamma_1 (d)$, where $d$ is a divisor of $N$. 
Alternatively, one can generate the full $\mathbb{Z}_N$ orbifold amplitude by considering orbits of the full untwisted sector with respect to the Hecke congruence subgroups $\Gamma_0 (d)$. However, in the cases of interest to string theory, it is often possible to generate the full $\Gamma_0 (d)$ orbit by using a single (untwisted) orbifold sector, thus simplifying the implementation of the orbit method. This happens, in particular, in the study of BPS-saturated contributions to gauge and gravitational threshold corrections in heterotic vacua, arising from the $\mathbb{Z}_N$ orbifold sectors, with  $N=2,3,4,6$. 

In dealing with integrals of the form \eqref{introcosetN} one has to properly address the problem of infrared (IR) divergences due to massless states propagating in the loop. Whereas in the original ${\rm SL} (2;\mathbb{Z})$-invariant description \eqref{introcoset} the IR divergence is associated to the unique boundary ({\em i.e.} the cusp at $\infty$) of the fundamental domain $\cF$, after partial unfolding it gets distributed among the various boundaries of $\cF_N$. These divergences necessitate an appropriate regularisation of the integral before  unfolding. As in \cite{MR656029,Angelantonj:2011br,Angelantonj:2012gw}, we introduce an explicit IR cut-off by truncating the ${\rm SL }(2;\mathbb{Z})$ fundamental domain $\cF$ to $\tau_2 <\cT$. This cut-off is then unambiguously extended to the other boundaries of $\cF_N$ after partial unfolding.

The outline of this work is as follows. In Section \ref{sec_orbifold} we review the general structure of closed strings on Abelian orbifolds with emphasis on the 
coset decompositions with respect to Hecke congruence subgroups. In Section \ref{sechecke} we list the main properties of  $\Gamma_0 (N)$ and extend the unfolding procedure to integrals over its truncated fundamental domain. In Section \ref{sec_rsz}, we discuss the Rankin-Selberg method for modular forms of $\Gamma_0(N)$ of moderate growth at all cusps, and apply it to evaluating integrals over $\cF_N$. We illustrate the procedure by explicitly computing integrals of freely acting orbifolds of $d$-dimensional Narain lattices, and express the result  in terms of level-$N$ generalisations of the constrained Epstein zeta series of \cite{Obers:1999um,Angelantonj:2011br,Angelantonj:2012gw}. In Section \ref{sec_np}, we introduce the absolutely convergent Niebur-Poincar\'e series associated to the Hecke congruence subgroups, and show how any weak almost holomorphic modular form can be represented by linear combinations of such series. We then explicitly compute the integral of a (shifted) Narain lattice times an arbitrary Niebur-Poincar\'e series and express the result in terms of a sum of Schwinger-like integrals associated to BPS-state contributions.
In Section \ref{sec_examples} we illustrate this method by studying a number of physically relevant examples of gauge and gravitational threshold corrections in 
freely-acting orbifolds and the free energy of an exotic class of two-dimensional superstring vacua. Appendix \ref{appendixKloosterman} collects useful formul\ae\ on   the Kloosterman-Selberg zeta function  for $\Gamma_0 (N)$ and the associated scattering matrices.  Finally, Appendix \ref{sec_comp} summarises useful facts on holomorphic modular forms of Hecke congruence subgroups.

\section{Generalities on one-loop amplitudes in closed string orbifolds}
\label{sec_orbifold}

In closed string theories compactified on orbifolds,  one-loop contributions to certain couplings in the low energy effective action typically take the form of an integral 
\begin{equation}
I = \int_{\cF} \de\mu \, \cA   \,, \qquad
\cA  =\sum_{h,g} \cA  \ar{h}{g} \, ,
\label{orbifoldintegral}
\end{equation}
where $\de\mu = \tau_2^{-2}\de\tau_1\, \de\tau_2$ is the ${\rm SL} (2; \mathbb{R})$ invariant measure on the Poincar\'e hyperbolic upper-half plane $\mathbb{H}$,  ${\cF} = \{ \tau \in \mathbb{H}\ | \ |\tau_1 | \leq \frac{1}{2}\,, |\tau|\ge 1\}$ is the standard fundamental domain of the full 
modular group $\Gamma = {\rm SL} (2 ; \mathbb{Z})$, corresponding to large reparametrisations of the two-dimensional world-sheet torus with Teichm\"uller parameter $\tau$, 
and $\cA$ is a modular form of weight 0
under $\Gamma$, whose explicit expression depends on the coupling under consideration. The sum over $h$ labels the various
twisted sectors and the sum over $g$ implements the orbifold projection onto invariant states. For
Abelian $\IZ_N$ orbifolds, the case of interest in this paper, both $h,g$ run over $0,\dots, N-1$. 
Although the untwisted unprojected sector of the orbifold $\cA  \ar{0}{0}$ is invariant under the full modular group $\Gamma$, this is not the case for each non-trivial orbifold block $ \cA  \ar{h}{g} $ with $(h,g)\neq (0,0)$ that is invariant only under a  congruence subgroup $\Gamma\ar{h}{g}\subset \Gamma$, {\em i.e.} 
a subgroup containing a  level-$M$ principal congruence subgroup
\begin{equation}
\label{GammaN}
\Gamma(M) = \left\{ \gamma= \left(\begin{array}{cc} a & b \\ c & d \end{array}\right)
 \in\Gamma \ \Bigg|\ a,d= 1\ {\rm mod} \ M\,, \ b,c=0\ {\rm mod}\ M \right\}\, .
\end{equation}
For $N$ prime, each $\Gamma\ar{h}{g}$ is conjugate to 
\begin{equation}
\label{Gamma1N}
\Gamma_1(N) = \left\{ \gamma = \left(\begin{array}{cc} a & b \\ c & d \end{array}\right) 
\in\Gamma \ \Bigg|\ a,d= 1\ {\rm mod} \ N\,, \ c=0\ {\rm mod}\ N \right\}\,,
\end{equation}
which has index $N^2-1$ with respect to the full modular group. Indeed, the $N^2-1$ non-trivial
orbifold blocks $ \cA  \ar{h}{g} $ form a single orbit of $\Gamma_1(N)\subset \Gamma$,
transforming as 
\be
\cA  \ar{h}{g} \Big\vert\gamma =\cA  \ar{cg+dh}{ag+bh}\,,
\label{modulartransformation}
\ee
where $|\gamma$ denotes the Petersson slash operator defined in Appendix \ref{sec_comp}, 
and can all be obtained from the block  $ \cA  \ar{0}{1}$, invariant under 
$\Gamma\ar{0}{1}=\Gamma_1(N)$. As a result, the sum
\be
\cA  =\sum_{h,g} \cA  \ar{h}{g}  = \mathcal{A}\ar{0}{0} + 
\sum_{\gamma \in  \Gamma_1(N) \backslash \Gamma}  \mathcal{A}\ar{0}{1} \vert\gamma 
\ee
is invariant under the full modular group, as expected. For our purposes, it will be convenient
to further collect the orbifold blocks into orbits of $\Gamma_0(N)\subset \Gamma$, where
$\Gamma_0(N)$ is the Hecke congruence subgroup of level $N$
\begin{equation}
\Gamma_0(N) = \left\{ \left(\begin{array}{cc} a & b \\ c & d \end{array}\right) 
\in\Gamma \ \Bigg|\ c=0\ {\rm mod}\ N \right\}\,, 
\label{Gamma0}
\end{equation}
which contains $\Gamma_1(N)$ as a normal subgroup such that 
$\Gamma_0(N)/\Gamma_1(N) = (\IZ/N\IZ)^*$, and has index
\begin{equation}
\nu_N=N \prod_{p|N}(1+p^{-1}) 
\label{fooindex}
\end{equation}
with respect to the full modular group $\Gamma$, where $p$ is a prime\footnote{In the following we shall always denote by $p$ a prime number.} divisor of $N$. 

The decomposition of the amplitude $\cA$ into orbits of the modular group is in general complicated, since
different sectors may be invariant under different level-$N_a$ congruence subgroups of $\Gamma$, where $N_a$ divides $N$. By collecting these contributions into orbits of $\Gamma_0(N_a)\backslash\Gamma$ one can thus write, in complete generality,
\begin{equation}
\cA  =  \sum_{N_a|N} \sum_{\gamma \in 
\Gamma_0(N_a) \backslash \Gamma}   \cA _{N_a} \Big\vert\gamma  \,, 
\label{decomp}
\end{equation}
where the untwisted unprojected contribution $\mathcal{A}\ar{0}{0}$ is  associated with 
$N_a=1$ in this sum.  
For $N$ prime the coset decomposition clearly simplifies since one single orbit with $N_a =N$ appears in the decomposition \eqref{decomp} (aside from the universal $N_a=1$ orbit) so that 
\be
\label{decorbprime}
\cA    = \mathcal{A}\ar{0}{0} + 
\sum_{\gamma \in  \Gamma_0(N) \backslash \varGamma}  \mathcal{\tilde A}\ar{0}{1}\Big\vert\gamma  \ ,\qquad {\rm with}\quad
\mathcal{\tilde A}\ar{0}{1} = \sum_{g=1}^{N-1} \mathcal{A}\ar{0}{g}\, .
\ee

In this paper, we shall consider in detail only the cases $N$ prime and $N=4$ and $N=6$,  since these are the only ones relevant for (half-)BPS-saturated amplitudes that
receive contributions only from the $\cN=2$ supersymmetric sectors, thus linked to the K3 orbifolds $T^4/\mathbb{Z}_N$ with $N=2,3,4,6$.
 For  $N=4, 6$ the decomposition \eqref{decomp} reads
\begin{equation}
\mathcal{A}= \cA  \ar{0}{0}
+  \sum_{\gamma \in  \Gamma_0(4) \backslash \Gamma}  ( \mathcal{A}\ar{0}{1}+\mathcal{A}\ar{0}{3} )\Big\vert\gamma 
+   \sum_{\gamma \in  \Gamma_0(2) \backslash \Gamma}  \mathcal{A}\ar{0}{2}\Big\vert\gamma \,,
\end{equation}
for $N=4$, and
\begin{equation}
\mathcal{A}=\cA  \ar{0}{0}
+  \sum_{\gamma \in  \Gamma_0(6) \backslash \Gamma}  (\mathcal{A}\ar{0}{1}+\mathcal{A}\ar{0}{5})\Big\vert\gamma 
+  \sum_{\gamma \in  \Gamma_0(3) \backslash \Gamma}  (\mathcal{A}\ar{0}{2}+ \mathcal{A}\ar{0}{4})\Big\vert\gamma 
+   \sum_{\gamma \in  \Gamma_0(2) \backslash \Gamma}  \mathcal{A}\ar{0}{3}\Big\vert\gamma \,,
\end{equation}
for $N=6$.

Having cast the integrand of \eqref{orbifoldintegral} into a finite sum \eqref{decomp} of orbits under
the Hecke congruence subgroups $\Gamma_0(N_a)$, one may use
\be
\label{FF0N}
{\cF}= \bigcup_{ \gamma\in \Gamma_0(N)\backslash \Gamma} \, \gamma\cdot {\cF}_N
\ee
to unfold the integration domain $\cF$  into the fundamental domain $\cF_{N_a}= \Gamma_0 (N_a) \backslash \mathbb{H}$ of 
$\Gamma_0(N_a)$, thus obtaining
\begin{equation}
I = \sum_{N_a|N} \, I_{N_a} \,,
\end{equation}
where 
\be
I_N = \int_{{\cF}_N} \de\mu\, \cA _N \,.
\label{orbifoldintegralN}
\ee
The next step consists then in devising methods to completely unfold $\cF_N$ into the usual strip $\cS =\{ 0\le \tau_2 <\infty \,,\ 0\le \tau_1 \le 1\}$, in a way that does not require a specific choice of Weyl chamber, that clearly spells out possible singularities of the amplitude at points of symmetry enhancement, and that maintains the perturbative symmetries of the vacuum manifest. As we shall see, special care is required for the proper definition of the modular integral and of its unfolding if IR divergences are present. We shall outline the general procedure in the next section, deferring the discussion of specific classes of integrals to Sections \ref{sec_rsz} and \ref{sec_np}.

\begin{figure}
\centerline{
\includegraphics[height=5cm]{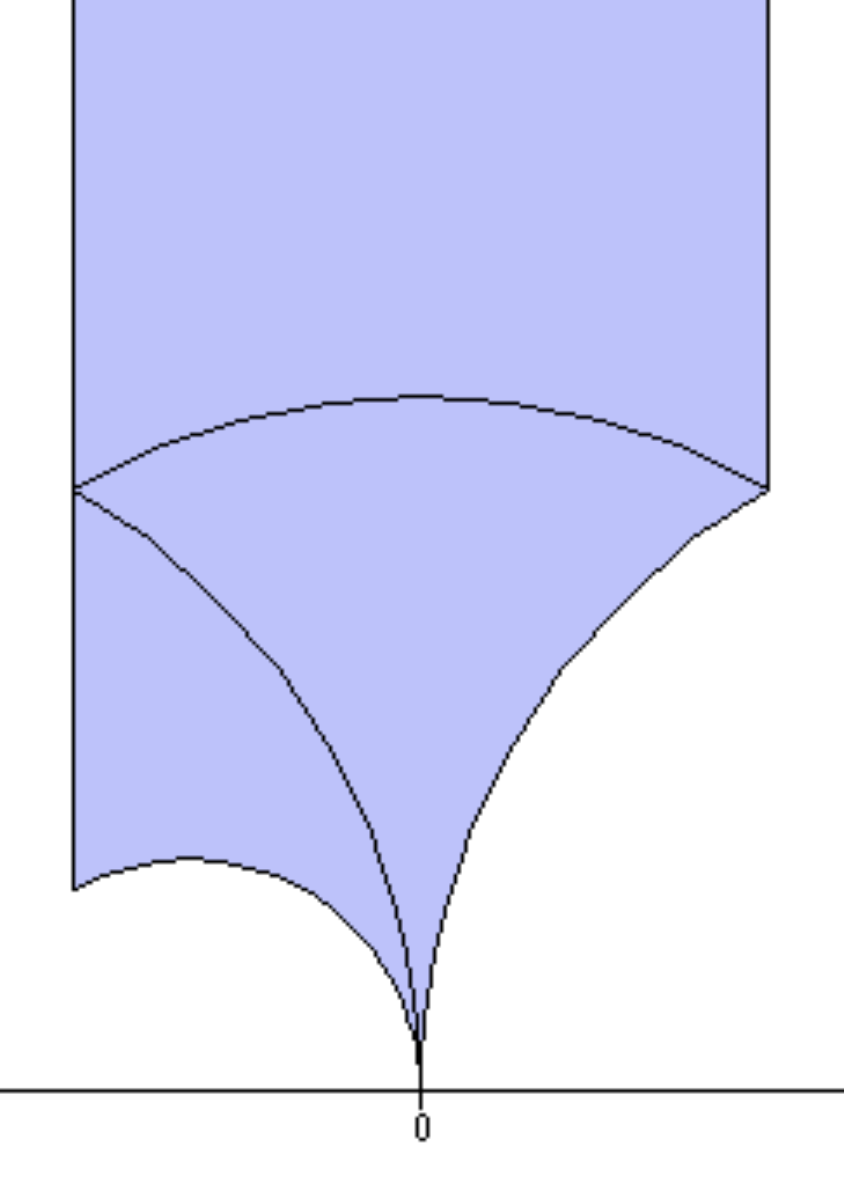}\hfill
\includegraphics[height=5cm]{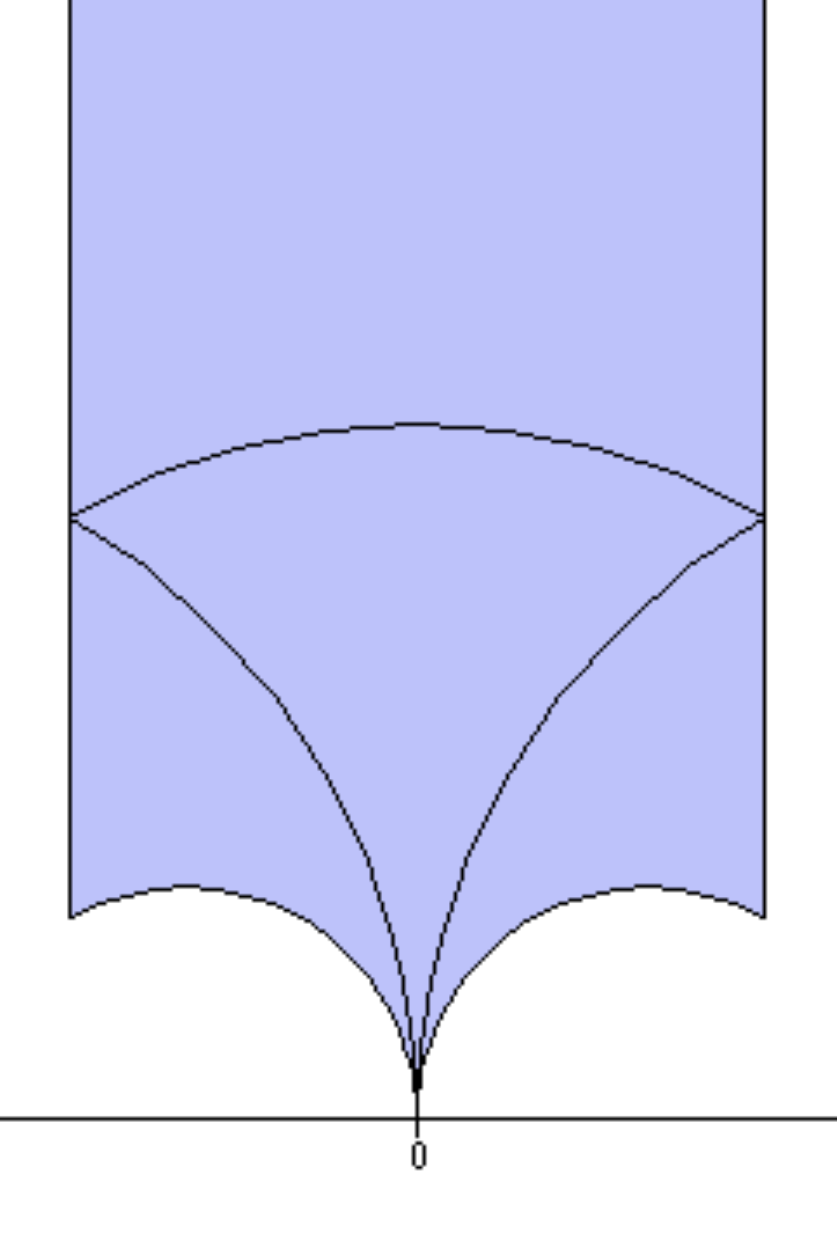}\hfill
\includegraphics[height=5cm]{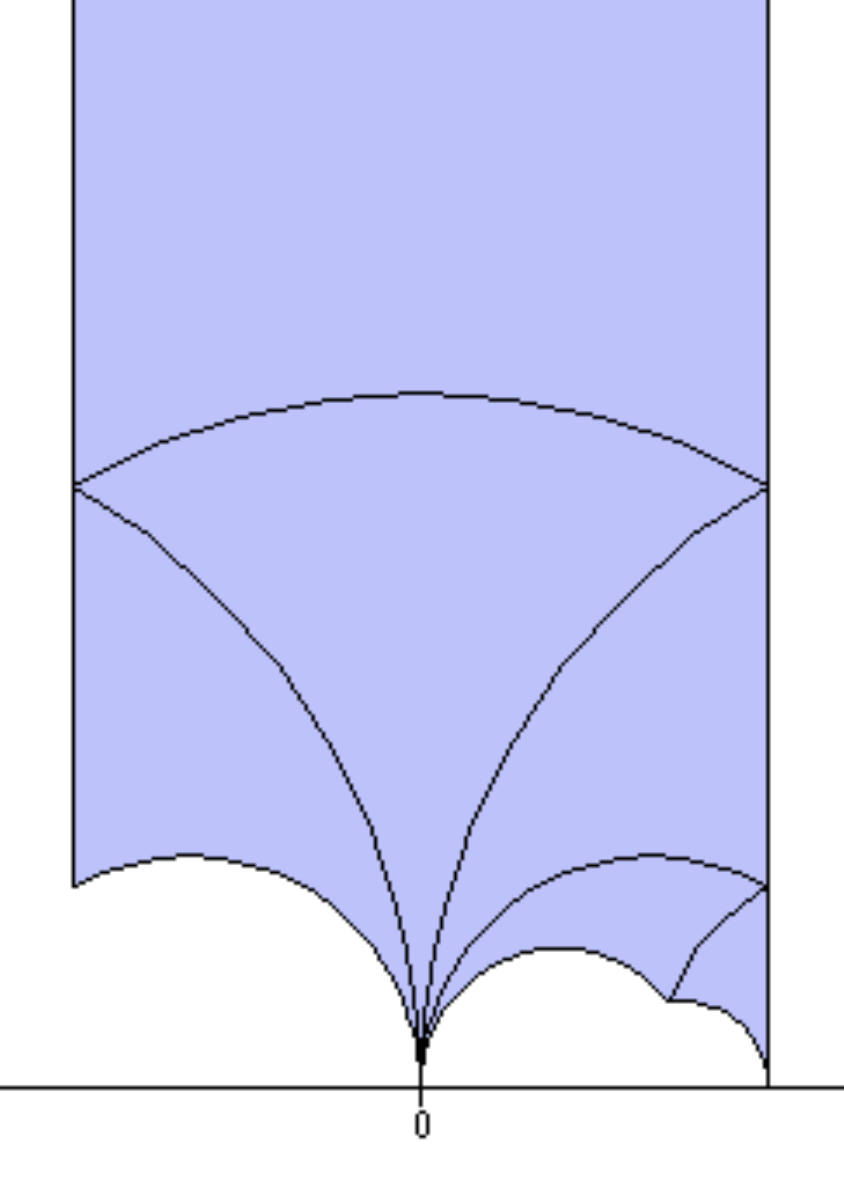}\hfill
\includegraphics[height=5cm]{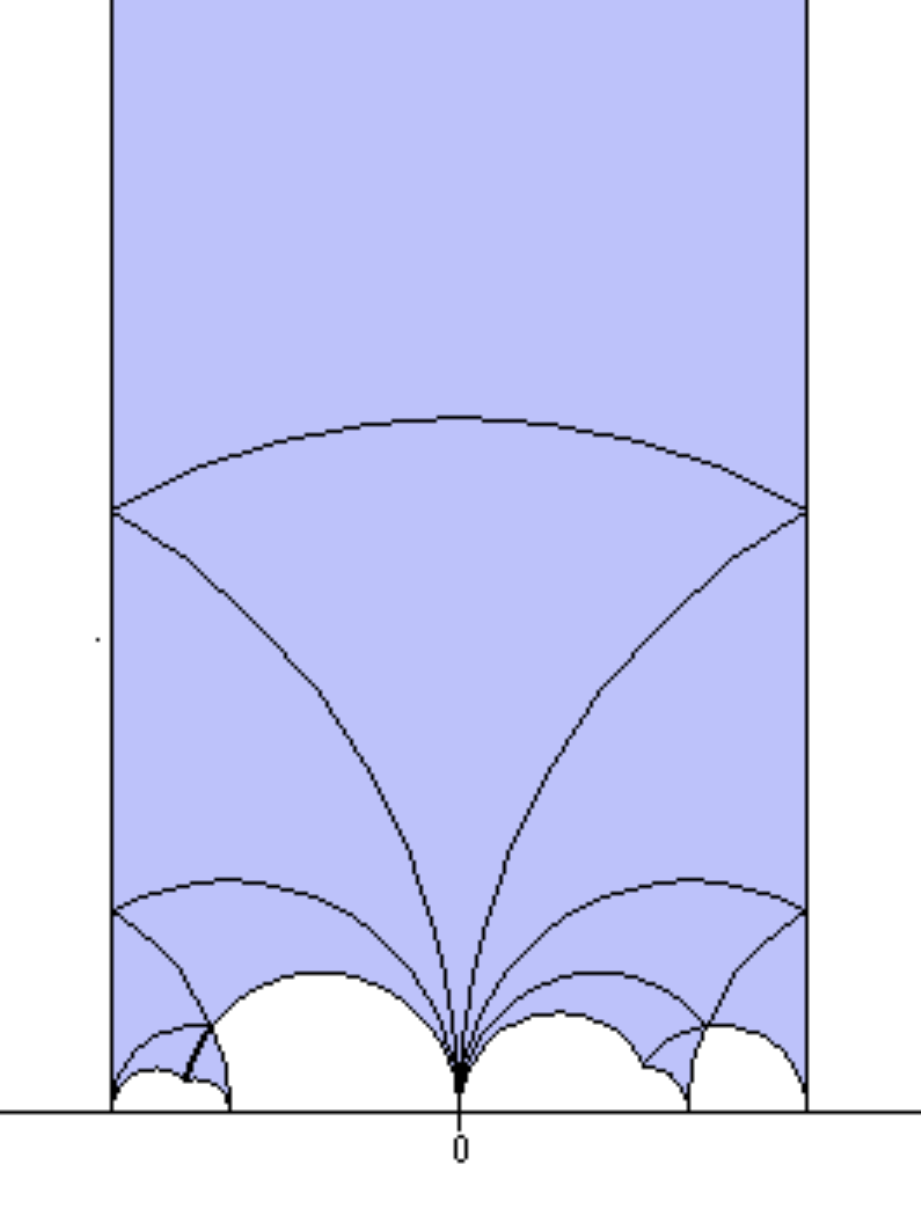}
}
\caption{Fundamental domains for the Hecke congruence subgroups $\Gamma_0(N)$
with $N=2,3,4,6$, respectively.}\label{figdomains}
\end{figure}

\section{The unfolding for Hecke congruence subgroups}\label{sechecke}

The goal of the following sections will be to extend the techniques introduced in  \cite{Angelantonj:2011br, Angelantonj:2012gw}, in the case of the full modular group $\Gamma$, to compute  integrals of the type \eqref{orbifoldintegralN}. These integrals typically develop IR singularities due to massless states running in the loop. 
Technically, these divergences arise at the boundary of the integration domain. In the case of the full modular group, the only boundary consists of the cusp at $\infty$, and a natural (modular invariant) way to deal with the IR singularity is to cut-off the fundamental domain at some large $\tau_2$. The main complication now is that the fundamental domain ${\cF}_{N}$ of the Hecke congruence subgroup $\Gamma_0 (N)$ has several boundaries associated to its inequivalent cusps, and each of them may give rise to divergences in the integral \eqref{orbifoldintegralN}. 
In principle, there are several options for regulating the integral but, as we shall argue, String Theory suggests a well-defined prescription. In the following, we shall first review some basic facts\footnote{Our discussion here follows \cite{shimura,iwaniec1,iwaniec2}, and the interested reader is referred to these references for more details.} about  $\Gamma_0 (N)$ and then we shall introduce the unfolding procedure and renormalisation prescription in some generality.

\subsection{Some properties of $\Gamma_0 (N)$ and its Poincar\'e series}

The Hecke congruence subgroup is defined by Eq. \eqref{Gamma0}. Similarly to the case of the full modular group, its fundamental domain $\cF_N = \Gamma_0 (N) \backslash \mathbb{H}$ can be compactified by adding $h$ inequivalent\footnote{Two points $z_1$ and $z_2$ are inequivalent if there is no element $\gamma \in \Gamma_0 (N)$ such that $z_2 = \gamma \cdot z_1$.} cusps, where
\begin{equation}
h=\sum_{d|N} \varphi ((d,N/d))\,, 
\end{equation}
$\varphi (n)$ being the Euler totient function. The cusps correspond to the rational points
\begin{equation}
\frac{u}{v}\qquad {\rm with}\quad v|N\,, \quad (u,v)=1\,, \quad u\ ({\rm mod}\ (v,N/v))\,.
\end{equation}
The cusps $1/N$ and $1$, equivalent to $\infty$ and $0$, respectively, always exist, and  are the only cusps for $N$ prime. In Figure \ref{figdomains} we have displayed the conventional choice for the fundamental domains of $\Gamma_0 (N)$ with $N=2,3,4,6$.

Each cusp $\mathfrak{a}$ can 
be obtained from the  cusp $\infty$ by acting with an element $\tau_{\mathfrak{a}}$
\begin{equation}
\tau_\mathfrak{a} = \left( \begin{array}{cc} u & * \\ v & * \end{array}\right)\in \Gamma_0(N)\backslash\Gamma\,, 
\end{equation}
so that $\mathfrak{a} =\tau_\mathfrak{a} \infty $. 
Clearly, $\Gamma_\mathfrak{a} = \tau_\mathfrak{a} \, \Gamma_\infty \, \tau_\mathfrak{a}^{-1}$ stabilises the cusp $\mathfrak{a}$ since $\Gamma_\infty$ is the stability group associated to $\infty$ and thus
\begin{equation}
\tau_\mathfrak{a}^{-1}\,\Gamma_\mathfrak{a}\, \tau_\mathfrak{a} =\left\{ \pm \left( \begin{array}{cc} 1 & b\, m_\mathfrak{a} \\ 0 & 1 \end{array}\right) \ \Big|\ b\in \mathbb{Z} \right\}\,.
\end{equation}
The positive integer $m_\mathfrak{a} = N/(N,v^2)$ is known as the width of the cusp $\mathfrak{a}$, and counts the number of copies of the fundamental domain of $\Gamma$ in 
\eqref{FF0N} whose boundary is the cusp $\mathfrak{a}$. One thus defines the {\em scaling matrix}
\begin{equation}
\label{defscalmat}
\sigma_\mathfrak{a} = \tau_\mathfrak{a}\, \rho_\mathfrak{a}\ ,\qquad
\rho_\mathfrak{a} = \left( \begin{array}{cc} \sqrt{m_\mathfrak{a}} & 0 \\ 0 & 1/\sqrt{m_\mathfrak{a}}\end{array}\right)\,,
\end{equation}
that reduces the width of each cusp to one. In general, $m_\infty =1$ and $m_0 = N$ and thus
\begin{equation}
\label{sigmaNprime}
\sigma_\infty =\left(\begin{array}{cc} 1& 0 \\ 0 & 1 \end{array}\right)\,,
\qquad
\sigma_0 = \left( \begin{array}{cc} 0 & 1/\sqrt{N} \\ -\sqrt{N} & 0 \end{array}\right)\, .
\end{equation}
The scaling matrices for the other cusps depend on the level $N$, and the relevant $\sigma_\mathfrak{a}$ for  $N=4,6$ can be found in Appendix \ref{appendixKloosterman}. 
 The scaling matrices  $\sigma_\mathfrak{a}$ can always be chosen such that 
\begin{equation}
\sigma_\mathfrak{a}\, \Gamma_0 (N) \, \sigma_\mathfrak{a}^{-1} = \Gamma_0 (N)\,,
\label{conjugation}
\end{equation}
a property that we shall exploit extensively for the unfolding procedure.

For any weight-$w$ modular form $G(\tau)$ of $\Gamma_0(N)$, one may define its Fourier expansion at any cusp $\mathfrak{b}$ as follows:  consider the modular form $G (\tau)\big\vert_w {\sigma_\mathfrak{b}}$, which is  periodic under $\tau\mapsto\tau+1$, and perform its usual Fourier expansion
\be
G (\tau)\Big\vert_w {\sigma_\mathfrak{b}}= j^{-w}_{\sigma_\mathfrak{b}} (\tau ) \, G ({\sigma_\mathfrak{b}}\tau ) =  \sum_{m\in\IZ} \tilde G_\mathfrak{b}(\tau_2 ; m)\, e^{2\pi\I m\tau_1}\,,
\label{modularfourier}
\ee
where
\be
j_\gamma (\tau ) = c \, \tau + d\,, \qquad {\rm for}\qquad \gamma = \begin{pmatrix} a & b \\ c & d \end{pmatrix} \in {\rm SL} (2;\mathbb{R}) \,.
\ee

The Poincar\'e series of $\Gamma_0 (N)$ with seed $f(\tau)$, periodic under $\tau \mapsto \tau +1$, and modular weight $w$, attached to the cusp $\mathfrak{a}$, is  defined by
\begin{equation}
F _\mathfrak{a} (\tau ) = \sum_{\gamma \in \Gamma_\mathfrak{a} \backslash \Gamma_0 (N)} f (\tau ) \Big\vert_w \sigma_\mathfrak{a}^{-1}\gamma
\,.\label{poincare}
\end{equation}
We shall require that the seed $f(\tau) \ll \tau_2^{1-w/2}$ as $\tau_2\to 0$ in order for \eqref{poincare} to be absolutely convergent so that it can be employed to unfold the fundamental domain $\cF_N$ as will be explained in the following subsection. Moreover, Eq. \eqref{conjugation} allows one to relate modular forms associated to different cusps, since
\begin{equation}
F_\mathfrak{a} (\tau ) = j_{\sigma_{\mathfrak{a}}^{-1}}^{-w}\, F_\infty ( \sigma_\mathfrak{a}^{-1}\tau ) = F_\infty (\tau) \big|_w\, \sigma_\mathfrak{a}^{-1}
\,.
\label{conjugaterelation}
\end{equation}
Finally, we denote its Fourier expansion around the cusp $\mathfrak{b}$ by
\begin{equation}
F_\mathfrak{a} (\tau ) \Big\vert_w \sigma_\mathfrak{b} = j^{-w}_{\sigma_\mathfrak{b}} (\tau ) \, F_\mathfrak{a} (\sigma_b \tau )=
\delta_\mathfrak{ab}\, f(\tau) +\sum_{m} \tilde F_\mathfrak{ab} (\tau_2 ; m )\, e^{2\pi \I m\tau_1}\,,
\end{equation}
where, here and henceforth, we isolate the contribution of the seed in the Fourier expansion of a Poincar\'e series.

\subsection{The unfolding procedure}\label{genunfolding}

We are now ready to proceed with the evaluation of the generic integral \eqref{orbifoldintegralN}, where the $\Gamma_0(N)$-invariant integrand $\mathcal{A}_N$  is assumed to factorise into the product of a weight-$w$  modular form $F_\mathfrak{a}$ times another modular form $G$
 with opposite weight so that the resulting integral \eqref{orbifoldintegralN}
  be well defined.  We shall assume that $F_\mathfrak{a}$ be represented by an absolutely convergent Poincar\'e series  as in \eqref{poincare}, since it will be used to unfold the fundamental domain. In our applications to BPS-saturated amplitudes, the role of $G$ will be  played by the (shifted) Narain lattice. Depending on the behaviour of $\cA_N$ at the various cusps, $F_\mathfrak{a}$ will be  given either by the non-holomorphic Eisenstein series or by the Niebur-Poincar\'e series. The latter provide a complete basis of weak almost holomorphic modular forms suitable for representing the (twisted) elliptic genus. In either case, the unfolding procedure, to be described below, leads to a rigorous definition of a renormalised integral. The procedure we shall present can be extended to more general classes of functions, however we shall not dwell upon these cases here.

Although at the level of the integral \eqref{orbifoldintegralN} one has several options to cope with the potential divergences, String Theory suggests a well-defined regularisation scheme.  Indeed, one-loop closed-string amplitudes arise in the form \eqref{orbifoldintegral}, where the integral runs over the fundamental domain of the full modular group and, thus, the only source of divergence in the original integral \eqref{orbifoldintegral} is from the cusp at $\infty$, and corresponds to the usual IR divergence of massless particles present in the spectrum. As in \cite{Angelantonj:2011br, Angelantonj:2012gw} this divergence can be regulated, consistently with modular invariance, by introducing a hard IR cut-off  $\tau_2\leq \cT$, thus replacing the ${\rm SL} (2;\mathbb{Z})$ fundamental domain $\cF$ by 
\begin{equation}
\mathcal{F}(\mathcal{T})=\mathcal{F}-\cS_\infty(\mathcal{T})\,,
\end{equation}
where $\cS_\infty(\mathcal{T})$ is the disk tangent to the cusp $\infty$, defined by $\cS_\infty(\mathcal{T}) = \bigr\{\tau\in\mathbb{H} \bigr| \tau_2\geq\mathcal{T} \bigr\}$.

The partial unfolding \eqref{FF0N} then lifts this regularisation scheme to the integral \eqref{orbifoldintegralN}, and hence the fundamental domain $\cF_N$ of $\Gamma_0 (N)$ is replaced by
\begin{equation}
\begin{split}
\mathcal{F}_N(\mathcal{T}) &\equiv \mathcal{F}_N - \bigcup\limits_{\mathfrak{b}}\ \tau_\mathfrak{b}\cdot \cS_\infty(\mathcal{T})
\\
&=\mathcal{F}_N - \bigcup\limits_{\mathfrak{b}}\ \tau_\mathfrak{b}\,\rho_\mathfrak{b}\,\rho_\mathfrak{b}^{-1}\cdot \cS_\infty(\mathcal{T}) 
\\
&=\mathcal{F}_N - \bigcup\limits_{\mathfrak{b}}\ \cS_\mathfrak{b}(\mathcal{T}) \,,
\end{split}
\label{circsubtract}
\end{equation}
where the small disks $\cS_\mathfrak{b}(\mathcal{T})\equiv \sigma_\mathfrak{b}\cdot \cS_\infty(m_\mathfrak{b}\mathcal{T})$ are excised around each cusp. Hence, one is led to the finite integral
\begin{equation}
I_\mathfrak{a}= \int_{\mathcal{F}_N(\mathcal{T})}\de\mu~ G(\tau)F_\mathfrak{a}(\tau)\,.
\end{equation}
Using the fact that $\sigma_\mathfrak{a}$ preserves $\Gamma_0(N)$ under conjugation \eqref{conjugation}, one may write $F_\mathfrak{a}(\tau)=j_{\sigma_\mathfrak{a}^{-1}}^{-w} (\tau )\, F_\infty(\sigma_\mathfrak{a}^{-1}\tau)$ and obtain
\begin{equation}
\begin{split}
I_\mathfrak{a}&=\int_{\sigma_\mathfrak{a}^{-1}\cdot\mathcal{F}_N(\mathcal{T})}\de\mu~G(\sigma_\mathfrak{a}\tau) \, j_{\sigma_\mathfrak{a}^{-1}}^{-w} (\sigma_\mathfrak{a}\tau )\,  F_\infty(\tau) 
\\
&= \int_{\sigma_\mathfrak{a}^{-1}\cdot\mathcal{F}_N(\mathcal{T})}\de\mu~G(\sigma_\mathfrak{a}\tau)\,  j_{\sigma_\mathfrak{a}}^{w} (\tau )\, 
F_\infty(\tau)\,,
\end{split}
\end{equation}
where we have made use of the identity $j_{AB} (\tau ) = j_A (B\tau) \, j_B (\tau)$, valid for any pair of matrices $A$,  $B$ in ${\rm SL} (2;\mathbb{R})$.
The Poincar\'e series $F_\infty (\tau)$ may now be employed in order to unfold the fundamental domain $\sigma_\mathfrak{a}^{-1}\cdot\mathcal{F}_N(\mathcal{T})$, since
\begin{equation}
\bigcup\limits_{\gamma\in\Gamma_\infty\backslash\Gamma_0(N)}\gamma\, \sigma_\mathfrak{a}^{-1}\mathcal{F}_N(\mathcal{T}) = \cS - 
\bigcup\limits_{\mathfrak{b}}\bigcup\limits_{\gamma \in\Gamma_\infty\backslash\Gamma_0(N)}\gamma\sigma_{\mathfrak{a}}^{-1}\cS_\mathfrak{b}(\mathcal{T})\,,
\end{equation}
with $\cS=\{\tau\in\mathbb{H}\,|\,-\frac{1}{2}\leq \tau_1\leq \frac{1}{2}\}$ being the half-infinite strip. As a result,
\begin{equation}
\begin{split}
I_\mathfrak{a} =& \int_{\cS}\de\mu\, G(\sigma_\mathfrak{a}\tau)\, j^w_{\sigma_\mathfrak{a} } (\tau )
\, f (\tau) -
\sum_{\mathfrak{b}}\int_{\cS_\infty(m_\mathfrak{b}\mathcal{T})}\de\mu~\tilde{G}_\mathfrak{b}(\tau_2;0)\left[\delta_{\mathfrak{ab}}f(\tau)+\tilde{F}_\mathfrak{ab}(\tau_2;0)\right]
\\
&-\sum_{\mathfrak{b}}\int_{\cS_\infty(m_\mathfrak{b}\mathcal{T})}\de\mu~G(\sigma_\mathfrak{b}\tau) \, j^w_{\sigma_\mathfrak{b} } (\tau )\left[F_\mathfrak{a}(\sigma_\mathfrak{b}\tau) \, j^{-w}_{\sigma_\mathfrak{b} } (\tau )
-\delta_{\mathfrak{ab}}f (\tau)-\tilde{F}_\mathfrak{ab}(\tau_2;0)\right] 
\\ 
& -\sum_{\mathfrak{b}}\int_{\cS_\infty(m_\mathfrak{b}\mathcal{T})}\de\mu~\left[G(\sigma_\mathfrak{b}\tau) \, j^w_{\sigma_\mathfrak{b} } (\tau ) - \tilde{G}_\mathfrak{b}(\tau_2;0)\right]\left[\delta_{\mathfrak{ab}}f (\tau)+\tilde{F}_\mathfrak{ab}(\tau_2;0)\right] 
\,,
\end{split}\label{generalresult}
\end{equation}
where $\tilde{F}_\mathfrak{ab}(\tau_2;0)$, $\tilde{G}_\mathfrak{b}(\tau_2;0)$ are the zero-frequency modes  of  the Fourier expansions of $F_\mathfrak{a}(\tau)$ and $G(\tau)$ at the cusp $\mathfrak{b}$, as defined in \eqref{poincare}, \eqref{modularfourier}, and 
 $f (\tau )$ is the seed of the Poincar\'e series $F_\mathfrak{a} (\tau )$. The advantage of the decomposition \eqref{generalresult} is that the dependence of $I_\mathfrak{a}$ on the cutoff $\mathcal{T}$ can be easily identified. Indeed, the integrands in the  second and third lines are exponentially suppressed as $\tau_2\rightarrow\infty$, while the second integral in the first line provides, in general, the non-trivial cut-off dependence and leads to the following natural definition of the renormalised integral
\begin{equation}
\begin{split}
{\rm R.N.}\int_{\mathcal{F}_N}\de\mu~ G(\tau)F_\mathfrak{a}(\tau)\equiv
	&\lim_{\mathcal{T}\rightarrow\infty}\Biggr[ \int_{\mathcal{F}_N(\mathcal{T})}\de\mu~G(\tau)F_\mathfrak{a}(\tau) \\
	& +\sum\limits_{\mathfrak{b}}\int_{\cS_\infty(m_\mathfrak{b}\mathcal{T})}\de\mu~\tilde{G}_\mathfrak{b}(\tau_2;0)\left[\delta_{\mathfrak{ab}}
	f (\tau)+\tilde{F}_\mathfrak{ab}(\tau_2;0)\right]\Biggr] \,,
	\end{split} \label{generalRNint}
\end{equation}
so that
\begin{equation}
{\rm R.N.}\int_{\mathcal{F}_N}\de\mu~ G(\tau)F_\mathfrak{a}(\tau)= \int_\cS \de\mu\, G(\sigma_\mathfrak{a}\tau)\, j^w_{\sigma_\mathfrak{a}} (\tau )\, 
f (\tau) \,.
\label{generalRNresult}
\end{equation}
Eqs. \eqref{generalresult}, \eqref{generalRNint} and \eqref{generalRNresult} provide the starting point for the methods that we shall introduce in the forthcoming sections.

\section{The Rankin-Selberg method and orbifolded lattices \label{sec_rsz}}

In this section we shall restrict our attention to the case where the integrand function
 $\cA _N$ is an automorphic  function under the Hecke congruence subgroup $\Gamma_0(N)$ with at most polynomial growth at all cusps. This case can be treated by the Rankin-Selberg method. Recall that the latter amounts to inserting a non-holomorphic Eisenstein series $E(\tau ,s)$ inside the integral, unfolding the integration domain against it for ${\rm Re}\, (s) \gg 1$, and analytically continuing the result to $s=1$ where $E(\tau,s)$ has a first order pole with constant residue. In Section \ref{sec_eisenstein} we shall give a general discussion on  non-holomorphic Eisenstein series for $\Gamma_0(N)$, and in Section \ref{sub_rsz} we shall expose the Rankin-Selberg method for congruence subgroups.

\subsection{Non-holomorphic Eisenstein series for $\Gamma_0(N)$}
\label{sec_eisenstein} 

The non-holomorphic, completed Eisenstein series associated to the cusp $\mathfrak{a}$ is defined by the sum over images
\begin{equation}
E^{\star}_\mathfrak{a} (\tau , s) =
\zeta^\star (2s)\, \sum_{\gamma \in \Gamma_\mathfrak{a}\backslash \Gamma_0(N)} \tau_2^s\, \Big\vert_0 \, {\sigma_\mathfrak{a}^{-1}\, \gamma}\ ,
\label{eisenstein}
\end{equation}
which is absolutely convergent for  ${\rm Re}\, (s) >1$. Here $\zeta^\star (s) = \pi^{-s/2} \varGamma (s/2 )\zeta (s)$ is the completed Riemann zeta function, whose introduction simplifies the functional relation below.
For $N=1$, the only cusp is  $\infty$ and Eq. \eqref{eisenstein} reduces to the usual
non-holomorphic Eisenstein series for the full modular group $\Gamma$. 

Since the seed $\tau^s_2$ is an eigenmode of the hyperbolic Laplacian $\Delta$, the Eisenstein series 
\eqref{eisenstein} satisfies
\be
\left[ \Delta + \tfrac12 s(1-s) \right]\, E^{\star}_\mathfrak{a} (\tau , s) = 0
\ee
for any cusp. The Fourier expansion of  $E^{\star}_\mathfrak{a}$ at the cusp $\mathfrak{b}$ is given by
\begin{equation}
E^{\star}_{\mathfrak{a}} (\sigma_\mathfrak{b} \tau, s ) =  e^{\star}_{\mathfrak{ab}} (\tau_2 ) + 2
\, \sum_{n\not= 0} \varphi_{\mathfrak{a}\mathfrak{b}} (n,s)\, 
 (|n|\tau_2)^{1/2}\, K_{s-1/2} (2\pi|n|\tau_2 ) \, e^{2\I\pi n\tau_1}\,,
\label{fourier}
\end{equation}
where
\begin{equation}
e^{\star}_{\mathfrak{ab}} (\tau_2 ) = \zeta^\star (2s) \left[ \delta_{\mathfrak{a}\mathfrak{b}}\, \tau_2^s + 
\pi^{1/2}\, \frac{\varGamma(s-\tfrac12)}{\Gamma(s)}
{\mathcal Z}_{\mathfrak{a}\mathfrak{b}} (0,0;s)
\tau_2^{1-s} \right]
\label{zeromode}
\end{equation}
is the zero-frequency mode, and
\begin{equation}
\begin{split}
\varphi_{\mathfrak{a}\mathfrak{b}} (n,s) &= \zeta  (2s) \, |n|^{s-1}\, 
{\mathcal Z}_{\mathfrak{a}\mathfrak{b}} (0,n;s)
\label{defvarphi}
\end{split}
\end{equation}
for non-vanishing frequencies.
Here ${\mathcal Z}_{\mathfrak{a}\mathfrak{b}}(0,n;s)$ is the Kloosterman-Selberg zeta function associated to the pair of cusps $\mathfrak{a}\mathfrak{b}$, defined in  \eqref{Zivgen}.
Selberg proved that $E^\star_\mathfrak{a} (\tau , s)$ has a meromorphic continuation to the whole $s$-plane, given by the Fourier expansion \eqref{fourier} itself.
Its simple poles are given by the poles of $e^\star_{\mathfrak{ab}}$. In particular, the point $s=1$ is a simple pole with constant residue 
\begin{equation}
{\rm Res}\ E^\star_\mathfrak{a} (\tau , s) \Big|_{s=1} = \frac{1}{2 \nu_N} = \frac{{\rm vol} (\cF )}{2\, {\rm vol} (\cF_N )}\,,
\label{ResEis}
\end{equation}
for any cusp $\mathfrak{a}$. Moreover, the Eisenstein series satisfy the functional equation
\begin{equation}
 E^{\star}_{\mathfrak{a}} (\tau , s) = \sum_\mathfrak{b}\varPhi_{\mathfrak{ab}}  (s) \, E^{\star}_{\mathfrak{b}}  (\tau , 1-s)\,, \label{functional}
\end{equation}
where the {\em scattering matrix} $\varPhi (s)$, with entries
\be
\varPhi_{\mathfrak{a}\mathfrak{b}}  (s) = \frac{\pi^{1/2} \, \varGamma(s-\tfrac12)\, \zeta^\star(2s)}
{\varGamma(s)\, \zeta^\star (2s-1) }
{\mathcal Z}_{\mathfrak{a}\mathfrak{b}} (0,0;s) \,,
\ee
satisfies $\varPhi (s) \, \varPhi (1-s) = \mathbb{1}$. 
The functional equation \eqref{functional} will play an important role in evaluating  the one-loop modular integrals in the next subsection.

For $N$ a square-free number one finds
\begin{equation}
\varPhi (s) = \bigotimes_{p|N} \cN_p (s)\,,\label{scattering}
\end{equation}
where
\begin{equation}
\cN_p (s) = \frac{1}{p^{2s}-1} \left( \begin{array}{cc} p-1 & p^s - p^{1-s}\\ p^s - p^{1-s} & p-1\end{array}\right)\,.\label{scattering2}
\end{equation}
In the more general cases where $N$ is not square-free the general expression is more complicated and the case for $N=4$ is given in Appendix \ref{appendixKloosterman}.

Notice that 
\be
E^{\star}_{\mathfrak{a}} (\tau,s) = E^{\star}_{\infty}(\sigma_\mathfrak{a}^{-1} \tau, s )\,,
\ee
as a result of \eqref{conjugation}, and one can always express the non-holomorphic Eisenstein series of $\Gamma_0 (N)$  as  linear combinations of the usual non-holomorphic Eisenstein series $E^{\star}$ of $\Gamma$ with suitably rescaled arguments. 
For instance, for $N$ prime, one may  show that
\begin{equation}
E^{\star}_\infty (\tau , s ) = \frac{N^s \, E^\star (N\tau , s) - E^\star (\tau , s)}{N^{2s} -1}\,,
\qquad 
E^{\star}_0 (\tau , s ) = \frac{N^s \, E^\star (\tau , s) - E^\star (N\tau , s)}{N^{2s} -1} \,. \label{relation}
\end{equation}
These relations allow one to straightforwardly compute the scattering matrix \eqref{scattering} using the functional equation of the ${\rm SL} (2;\mathbb{Z})$ Eisenstein series, $E^\star (\tau,s) = E^\star (\tau , 1-s)$.

Eq. \eqref{relation} and the first Kronecker limit formula for the standard
Eisenstein series allow one to extract similar limit formul\ae\ for the $E^\star_\mathfrak{a} (\tau ,s)$ Eisenstein series. In particular, for $N$ prime one finds
\begin{equation}
\begin{split}
E^{\star}_\infty (\tau ,s) =& \frac{1}{2 (N+1)\, (s-1)}
\\
&-\frac{1}{2 (N^2 -1)}\,\log \left[ (4\pi)^{N-1} \, e^{(1-N)\gamma }\, N^{2 N^2 /(N+1)} \, \tau_2^{N-1}\, \left| \frac{[\eta (N\tau )]^N}{\eta (\tau )} \right|^4 \right] +\ldots \,,
\\
E^{\star}_0 ( \tau , s) =& \frac{1}{2 (N+1)\, (s-1)}
\\
&-\frac{1}{2 (N^2 -1)}\,\log \left[ (4\pi)^{N-1} \, e^{(1-N)\gamma }\, N^{( N^2-2N-1) /(N+1)} \, \tau_2^{N-1}\, \left| \frac{[\eta (\tau )]^N}{\eta (N\tau )} \right|^4 \right] +\ldots \,.
\end{split} \label{kronecker}
\end{equation}
For $N=4,6$ similar expressions can be derived using Eqs. \eqref{relation4} and \eqref{relation6}.

\subsection{The Rankin-Selberg method for Hecke congruence subgroups}\label{sub_rsz}

Having defined the Eisenstein series for $\Gamma_0(N)$, we now turn to the evaluation of the integral
\eqref{orbifoldintegralN} by following the unfolding procedure outlined in Section \ref{genunfolding} (see also \cite{gupta}).  As already mentioned, in this case the integrand $\cA _N$ is a (not necessarily holomorphic) automorphic function of $\Gamma_0(N)$, with  at most  power-like growth at each cusp\footnote{The techniques we are going to outline in this section can be actually extended to cases of more general growth at the cusp ($\tau_2^\alpha (\log \tau_2)^n$). However, we shall limit ourselves to power-like behaviour since this is the only one of interest in String Theory.}
\begin{equation}
\cA _N (\sigma_\mathfrak{a} \, \tau ) \sim \psi_{\mathfrak a} (\tau_2 ) + O(\tau_2^{-M} )
\,, \qquad \forall M>0\,,
\end{equation}
as $\tau_2 \to \infty$, where
\begin{equation}
\psi_\mathfrak{a} (\tau_2 ) =\sum_i c_{\mathfrak{a},i}\, \tau_2^{\alpha_{\mathfrak{a},i}}\,, \qquad c_{\mathfrak{a},i}\,, \ \alpha_{\mathfrak{a},i}\in\mathbb{C}\,.
\end{equation}

To evaluate \eqref{orbifoldintegralN},  we apply the Rankin-Selberg method and consider,
for a generic cusp $\mathfrak{a}$, the integral
\begin{equation}
I_N (s; \cT) = \int_{\cF_N (\cT )} \de\mu\, \cA _N (\tau ) \, E^{\star}_\mathfrak{a} (\tau , s ) \,, \label{RSintegral}
\end{equation}
where the Poincar\'e series $F_\mathfrak{a}$ of Section \ref{genunfolding} is now replaced by $E^{\star}_\mathfrak{a} (\tau , s )$, that converges absolutely for ${\rm Re}\, (s) >1$, and is thus suited for unfolding. 

Using Eq. \eqref{generalresult} adapted to the case at hand, one obtains
\begin{equation}
\begin{split}
\mathcal{R}^\star_\mathfrak{a} (\cA_N , s)  =&
\int_{\cF_N  (\cT)} \de\mu\, \cA_N(\tau) \, E^{\star}_\mathfrak{a} (\tau , s ) 
\\
&+ \sum_\mathfrak{b} \int_{\mathcal{S}_\infty (m_\mathfrak{b}\mathcal{T})} \de\mu 
\left[ \cA_N  (\sigma_\mathfrak{b} \tau )\,  E^{\star}_\mathfrak{a} (\sigma_\mathfrak{b} \tau, s ) 
- \psi_\mathfrak{b} (\tau_2 ) \, e^{\star}_\mathfrak{ab} (\tau_2) \right]
\\ 
&- \zeta^\star (2 s) \, h_{\mathfrak{a}} (m_\mathfrak{a}{\mathcal T},s) -  \zeta^\star (2s-1)\sum_\mathfrak{b} \varPhi_{\mathfrak{a}\mathfrak{b} } (s) \, h_\mathfrak{b} (m_\mathfrak{b}\mathcal{T} , 1-s) \,,
\end{split}\label{RSintegral2}
\end{equation}
where $e^{\star}_\mathfrak{ab} (\tau_2 ) $ is the zero-mode \eqref{zeromode}, 
\begin{equation}
h_\mathfrak{a} ({\mathcal T},s) =\int_0^\mathcal{T} d y\, y^{s-2}\, \psi_\mathfrak{a} (y) = \sum_i c_{\mathfrak{a}, i}\, \frac{\mathcal{T}^{s+\alpha_{\mathfrak{a},i}-1}}{s+\alpha_{\mathfrak{a},i}-1}
\end{equation}
 and 
 \begin{equation}
\mathcal{R}^\star_\mathfrak{a} (\cA _N, s)  
= \zeta^\star (2s ) \int_0^\infty dy\, y^{s-2} \left [a_\mathfrak{a} (y) -\psi_\mathfrak{a} (y) \right]
\end{equation}
is the Rankin-Selberg transform associated to the cup $\mathfrak{a}$, with
\begin{equation}
a_\mathfrak{a} (\tau_2 ) = \int_0^1 d\tau_1 \, \cA_N  (\sigma_\mathfrak{a}\tau )  \,.
\end{equation}
Eq. \eqref{RSintegral2} defines the meromorphic continuation to the whole complex $s$-plane of the Rankin-Selberg transform, since the second integral on the {\em r.h.s.} is clearly finite in the ${\mathcal T} \to\infty$ limit and defines an entire function of $s$, while the remaining terms are meromorphic functions of $s$. As a result, $\cR^\star_\mathfrak{a}$  has simple poles\footnote{The poles are simple as long as $\alpha_{\mathfrak{a},i} \not= 1$, otherwise a double pole develops at $s=1$.} at $s=0,\, 1,\, \alpha_{\mathfrak{a},i},1-\alpha_{\mathfrak{a},i}$, and inherits the functional relation \eqref{functional}
\begin{equation}
{\mathcal R} ^\star_{\mathfrak{a}} (\cA_N ,s ) = \sum_\mathfrak{b} \varPhi_{\mathfrak{ab}} (s) \, 
 {\mathcal R}^\star_{\mathfrak{b}} (\cA_N , 1-s )\,. \label{RSfunctional}
\end{equation}

The residue of Eq. \eqref{RSintegral2} at $s=1$ is particularly useful, since it allows one to make contact with the integral \eqref{orbifoldintegralN}. 
Indeed, upon defining the renormalised integral 
\begin{equation}
{\rm R.N.} \int_{{\cF}_N}\de\mu\, \cA_N  
= \lim_{\mathcal{T}\to\infty} \left[ \int_{\cF_N(\cT )}
\de\mu\, \cA_N   -  \sum_\mathfrak{b} \hat\psi_{\mathfrak{b}} (m_\mathfrak{b}\mathcal T) 
\right]\,,
\end{equation}
where 
\begin{equation}
\hat\psi_\mathfrak{a} (y) =\sum_{\alpha_{\mathfrak{a},i}\not=1} c_{\mathfrak{a},i} \frac{y^{\alpha_{\mathfrak{a},i}-1}}{\alpha_{\mathfrak{a},i}-1}+\sum_{\alpha_{\mathfrak{a},i}=1} c_{\mathfrak{a},i}\, \log\, y
\end{equation}
is the anti-derivative of $\psi_\mathfrak{a} (y)$ on the hyperbolic plane, one obtains
\begin{equation}
\begin{split}
{\rm R.N.} \int_{{\cF}_N}\,\de\mu\, \cA_N  =& \,2\nu_N \,
{\rm Res}\, \Biggl[ \mathcal{R}^\star_\mathfrak{a} (\cA_N ,s) +\zeta^\star (2s) \, h_\mathfrak{a} (m_\mathfrak{a}\mathcal{T},s) 
\\&
+\zeta^\star (2s-1) \sum_\mathfrak{b}\varPhi_{\mathfrak{ab}} (s)\, h_\mathfrak{b} (m_\mathfrak{b}\mathcal{T},1-s)
\Biggr]_{s=1}
- \sum_\mathfrak{a} \hat\psi_\mathfrak{b} (m_\mathfrak{b}\mathcal{T})\,.
\end{split}
\label{RSrenormalised}
\end{equation}
As a trivial example, taking $\cA_N $ to be the unit function, the Rankin-Selberg transforms vanish, and one recovers the expected volume $\frac{1}{3}\pi \nu_N$ of the fundamental domain of $\Gamma_0(N)$.

\subsection{Shifted lattice integrals and Epstein zeta functions}

We now have all the necessary ingredients to compute the one-loop integral of shifted Narain lattices, that arise, for instance, in one-loop threshold corrections to
low-energy couplings of heterotic and type II superstrings, possibly with partial supersymmetry breaking. Let us consider the case of a $\mathbb{Z}_N$ freely-acting orbifold of a $d$-dimensional lattice with background metric $G_{ij}$ and Kalb-Ramond field $B_{ij}$. For simplicity, we restrict our analysis to the case of $N$ prime, though generalisation to the case where $N$ is not prime is straightforward. The resulting lattice partition function reads
\begin{equation}
\cA  = \frac{1}{N}\, \sum_{h,g=0}^{N-1} \varGamma_{(d,d)} \ar{h}{g} (G,B;\tau)\,, \label{Narainpartition}
\end{equation}
where
\begin{equation}
\varGamma_{(d,d)} \ar{h}{g}= \tau_2^{d/2}\, \sum_{\vec p\in \mathbb{Z}^{2d}} e^{2\I\pi g\, \vec \lambda\cdot \vec p}\, e^{\I \pi \tau_1 (\vec p +h\vec \lambda)^2} \, e^{-\pi\tau_2 {\mathcal M}^2 (h)}\,. \label{Narainlattice}
\end{equation}
Here $\vec p = (\vec m \,,\,\vec n)$ is a $2d$-dimensional integral vector encoding the momentum and winding quantum numbers, and $\vec\lambda = (\vec \lambda_1 \,,\, \vec \lambda_2 )$ is a constant 
$2d$-dimensional  vector acting as a shift along the momenta ($\vec \lambda _2$) and/or windings ($\vec\lambda_1$). Its entries can be taken to be $k/N$, with $k\in \mathbb{Z}_N$, and it must satisfy the constraint
\begin{equation}
N\, \vec \lambda \cdot \vec\lambda = 0 \ {\rm mod}\ 2 \label{constraint}
\end{equation}
in order to ensure modular invariance. The scalar product is defined with respect to the ${\rm O}(d,d)$-invariant  metric $\varOmega$,
\begin{equation}
\vec v \cdot \vec w \equiv \vec v ^T \, \varOmega \, \vec w = \left( \begin{array}{cc} \vec v_1 & \vec v_2 \end{array}\right) \left( \begin{array}{cc} 0 &\mathbb{1} \\ 
\mathbb{1} & 0 \end{array}\right) \left( \begin{array}{c} \vec w_1 \\ \vec w_2 \end{array} \right) \,,
\end{equation}
and ${\vec v} ^2 \equiv \vec v \cdot \vec v$. The BPS squared-mass depends on the shift vector
\begin{equation}
{\mathcal M}^2 (h) = (\vec p + h \, \vec \lambda)^T \, M^2 \, (\vec p + h \, \vec \lambda) \,,
\end{equation}
and on the background moduli through the matrix
\begin{equation}
M^2 = \left( \begin{array}{cc} G^{-1} & G^{-1} B \\ - B G^{-1} & G - B G^{-1} B \end{array}\right) \,.
\end{equation}

Although \eqref{Narainpartition} can be shown to correspond to a (unshifted) Narain lattice partition function on a different background \cite{Gregori:1997hi}, and can thus be integrated following 
\cite{Angelantonj:2011br}, it will be instructive to use instead the coset decomposition 
\begin{equation}
\frac{1}{N}\, \sum_{h,g=0}^{N-1} \varGamma_{(d,d)} \ar{h}{g}
= \frac{1}{N} \, \varGamma_{(d,d)} \ar{0}{0}   +\frac{1}{N}
\sum_{\gamma \in \Gamma_0(N)\backslash \Gamma } \, \sum_{g=1}^{N-1}
\varGamma_{(d,d)} \ar{0}{g}\Big\vert\gamma  \,.
\end{equation}

Therefore, the one-loop modular integral of the orbifolded Narain lattice decomposes as 
\begin{equation}
 \int_{\cF}\de\mu \,\frac{1}{N} \sum_{h,g=0}^{N-1} \varGamma_{(d,d)} \ar{h}{g}
 =  \int_{\cF}\de\mu \, \frac{1}{N}
\varGamma_{(d,d)}\ar{0}{0}+ \int_{\cF_N }\de\mu \, \frac{1}{N}\sum_{g=1}^{N-1}
\varGamma_{(d,d)} \ar{0}{g}  \,, 
\label{onelooporbNarain}
\end{equation}
and one can apply the procedure exposed in the previous subsection to compute the latter, since the 
lattice $\varGamma_{(d,d)} \ar{0}{g}$ has at most polynomial growth at the two cusps\footnote{Actually, for a non-vanishing shift vector $\vec\lambda$, it has the familiar behaviour $\tau_2^{d/2}$ as  $\tau_2 \to \infty$ whereas it is exponentially suppressed at $0$.}.
As a result, the Rankin-Selberg transforms read
\begin{equation}
\begin{split}
\mathcal{R}^\star_\infty (\varGamma_{(d,d)} \ar{0}{g}, s) &= \zeta^\star (2 s) \int_0^\infty d t \, t^{s+d/2-2} \sum_{{\vec p \in \mathbb{Z}^{2d}\atop \vec p \not= 0}} e^{2\I \pi g \vec\lambda \cdot \vec p}\, e^{-\pi t \, {\mathcal M}^2 (0)} \, \delta (\vec p^2 )
\\
&= {\mathcal E}^\star_d \ar{0}{g}(G,B;s+\tfrac{d}{2}-1 )
\end{split}
\end{equation}
and
\begin{equation}
\begin{split}
\mathcal{R}^\star_0 (\varGamma_{(d,d)} \ar{0}{g}, s) &= \zeta^\star (2 s) N^{d/2} \int_0^\infty d t \, t^{s+d/2-2} \sum_{\vec p \in \mathbb{Z}^{2d}} e^{-\pi N\,t \, {\mathcal M}^2 (1)} \, \delta ((\vec p +g\vec\lambda )^2 )
\\
&= N^{1-s}\,  {\mathcal E}^\star_d \left[ {\textstyle{g\atop 0}}\right] (G,B;s+\tfrac{d}{2}-1 )
\end{split}
\end{equation}
where we have defined the completed constrained Epstein zeta function with characteristics
\begin{equation}
\begin{split}
{\mathcal E}^\star_d \ar{h}{g} (G,B;s) &\equiv \pi^{-s}\, \varGamma (s) \, \zeta^\star (2s-d+2) \, 
{\mathcal E}_d \ar{h}{g}(G,B;s)
\\
&\equiv \pi^{-s}\, \varGamma (s) \, \zeta^\star (2s-d+2) \, {\sum_{{\vec p \in \mathbb{Z}^{2d}}\atop \vec p \not=0\ {\rm if}\ h=0}}
e^{2\I \pi g \vec\lambda \cdot \vec p}\, [{\mathcal M}^{2} (h)]^{-s} \, \delta ((\vec p +h\vec\lambda )^2 ) \,.
\end{split}
\end{equation}
Similar expressions clearly hold for $N$ non-prime and can be easily worked out following the discussion in Section \ref{sec_orbifold}.

Combining the results in \cite{Angelantonj:2011br} with those in the previous subsection, the renormalised one-loop modular integral \eqref{onelooporbNarain} then reads
\begin{equation}
\begin{split}
{\rm R.N.} \int_{\cF}\de\mu \frac{1}{N}\, \sum_{h,g=0}^{N-1} 
\varGamma_{(d,d)} \ar{h}{g}(G,B;\tau) =& \frac{2}{N} \,
{\rm Res}\, {\mathcal E}^\star_d\ar{0}{0}(G,B;s+\tfrac{d}{2}-1 ) \Big|_{s=1}
\\
&+ \frac{2 (N+1)}{N}\sum_{g=1}^{N-1} \,{\rm Res}\, 
{\mathcal E}^\star_d \ar{0}{g}(G,B;s+\tfrac{d}{2}-1 ) \Big|_{s=1} \,.
\end{split}
\end{equation}
Extracting the residue of the completed Epstein zeta functions $\cE^\star_d$ can be a non-trivial task, since the simple pole is associated to the Epstein zeta function $\cE_d$ itself, and its analytic properties are not always under control. However, similarly to \cite{Angelantonj:2011br}, one can make use of the functional relation \eqref{RSfunctional} to simplify the task of extracting the residue, since now the simple pole is associated to the overall Euler $\varGamma$-function, while the Epstein zeta function is analytic at $s=1$. In fact, Eq. \eqref{RSfunctional} translates into the following functional relation for the completed Epstein zeta functions with characteristics\footnote{To lighten the notation, we omit the explicit dependence of the lattice and of the Epstein zeta-functions on the geometric moduli, unless  needed.}
\begin{equation}
{\mathcal E}^\star_d \ar{0}{g}(s+\tfrac{d}{2}-1) =  \varPhi_{\infty\, \infty} (s) \, {\mathcal E}^\star_d \ar{0}{g}(\tfrac{d}{2}-s) + \varPhi_{\infty \, 0} (s) \, N^{1-s}\, {\mathcal E}^\star_d \ar{g}{0}(\tfrac{d}{2}-s) \,. \label{epsteinfunctional}
\end{equation}
As a result,  one finds
\begin{equation}
\begin{split}
{\rm R.N.} \int_{\cF}\de\mu \frac{1}{N}\, \sum_{h,g=0}^{N-1} 
\varGamma_{(d,d)} \ar{h}{g}(\tau) =& \frac{\pi^{1-d/2}\, \varGamma (\frac{d}{2}-1) }{N} \left[ {\mathcal E}_d \ar{0}{0} (\tfrac{d}{2}-1 ) \right.
\\
&+ \left.  \sum_{g=1}^{N-1} \left( {\mathcal E}_d \ar{0}{g}(\tfrac{d}{2}-1 ) + 
{\mathcal E}_d \ar{g}{0}(\tfrac{d}{2}-1 ) \right)\right] \,,
\end{split}
\end{equation}
valid for any dimension $d\not= 2$, at any point in the Narain moduli space and for any choice of shift vector $\vec\lambda$. Special attention is required in the two-dimensional case, since $\alpha_{\infty}=d/2=1$, and the pole at $s=1$ is now double. This case will be discussed in detail in the next subsection. 

Let us conclude this general discussion with a comment on the symmetries of the Epstein zeta functions with characteristics. Although, upon a suitable redefinition of the background fields, the orbifolded Narain lattice partition function \eqref{Narainpartition} is invariant under the full $T$-duality group ${\rm O} (d,d;\mathbb{Z})$, this is not the case for the individual contributions\footnote{Aside for the contribution $(h,g)=(0,0)$ that corresponds to the original Narain lattice.} $\varGamma_{(d,d)} \ar{h}{g}$, that are  invariant only  with respect to the subgroup ${\rm O}_{\vec\lambda} (d,d;\mathbb{Z} )$ which fixes the shift vector $\vec \lambda$ modulo $\mathbb{Z}^{2d}$. 

\subsection{Lower-dimensional lattices}

To illustrate the procedure, let us study in detail the lower-dimensional cases. For $d=1$ the only possible choice\footnote{up to the inversion of the compactification radius} of the vector compatible with the constraint \eqref{constraint} is $\vec\lambda = (0,1/N)$. In this case, corresponding to a momentum shift, it is easier to compute the Rankin-Selberg transform associated to the cusp at 0, ${\mathcal R}^\star_0 (
\varGamma_{(1,1)}\ar{0}{g} ;s) = N^{1-s} {\mathcal E}^\star_1\ar{g}{0}(R;s-1/2)$, 
since the constraint $m (n+g/N)=0$ has the unique solution $m=0$. One thus finds
\begin{equation}
{\mathcal E}^\star_1 \ar{g}{0}(R; s-\tfrac{1}{2}) = 
\pi^{1/2-s}\varGamma (s-\tfrac{1}{2}) \zeta^\star (2s) R^{1-2s}
\Bigl[ \zeta (2s-1;g/N)+\zeta (2s-1;1-g/N)\Bigr]\,. \label{onedimepstein}
\end{equation}
We remind here that the Hurwitz zeta function $\zeta (s;a)$ has a simple pole at $s=1$ whose residue is independent of $a$ and equals 1.
Therefore, computing the residue at $s=1$ of Eq. \eqref{onedimepstein} yields 
\begin{equation}
\int_{\cF_N} \de\mu \, \sum_{g=1}^{N-1}\, \varGamma_{(1,1)}\ar{0}{g} (R;\tau) = \frac{\pi \,(N^2-1) }{3 R}\,,
\end{equation}
where again we have restricted our analysis to the case $N$ prime. 
This result is compatible with  
\begin{equation}
 \int_{\cF}\de\mu\, \frac{1}{N}\sum_{h,g=0}^{N-1}\, \varGamma_{(1,1)}\ar{h}{g}(R;\tau) = 
\frac{\pi}{3} \left( \frac{R}{N} + \frac{N}{R}\right)\,,
 \end{equation}
in accordance with the fact that the $\mathbb{Z}_N$ momentum-shift orbifold of the one-dimensional Narain lattice has the net effect of diving by $N$ the radius of the compactification circle, $R\to R/N$.  
 
As a further example, let us consider the $\mathbb{Z}_2$ shift orbifold of a two-dimensional lattice with complex structure $U$ and K\"ahler form $T$. In this case, the constraint \eqref{constraint} admits several inequivalent solutions which correspond to different (discrete) marginal deformations of the lattice. In the following we shall treat explicitly the case $\vec\lambda = (0,0;\frac{1}{2},0)$, since the other choices can be worked out in a similar fashion. In this case, the evaluation of the integral 
\begin{equation}
{\rm R.N.}\int_{\cF_2}\de\mu \,\varGamma_{(2,2)} \ar{0}{1} (U,T) = 6 \, {\rm Res}\, \mathcal{E}^\star_2 \ar{0}{1} (U,T; s) \Big|_{s=1} +\log\left( 2^{14/3}\,\pi\,e^{-\gamma}\right)
\end{equation}
requires special care since, for $d=2$, the Rankin-Selberg transform has a double pole at $s=1$ and the functional equation \eqref{epsteinfunctional} is not sufficient to explicitly extract the residue. However, in this case one can explicitly solve the constraint $m_1 n^1 + m_2 n^2 =0$ and express
the constrained Epstein zeta function $\mathcal{E}^\star_2 \ar{0}{1} (U,T;s)$ in terms of Eisenstein series whose analytic properties are 
well-known and can be used to cast the result in terms of known functions. The set of solutions to the Diophantine equation is
\begin{equation}
\Biggr\{
\begin{split}
& S_1 ~:~ (m_1,m_2;n^1,n^2) = (m_1,m_2;0,0)\,, \quad m_1,m_2\in\mathbb{Z} \,,
\\
& S_2 ~:~ (m_1,m_2;n^1,n^2) = (c \tilde{m}_1,c \tilde{m}_2;-d\tilde{m}_2,d \tilde{m}_1)\,, \quad (\tilde{m}_1,\tilde{m}_2)=1 ~,~ c\in\mathbb{Z}~,~d\geq 1 \,,
\end{split}
\end{equation}
and after some algebra one gets
\begin{equation}
\mathcal{E}^\star_2 \left[ {\textstyle{0\atop 1}}\right] (U,T;s) = 4\left[ E^\star (\tfrac{1}{2}T,s) \, E^\star_0 (U,s) + E^\star (T,s) \, E^\star_\infty (U,s) \right] - 2 E^\star (T,s)\, E^\star (U,s)\,,
\end{equation}
where $E^\star (z,s)$ is the completed ${\rm SL} (2;\mathbb{Z})$ Eisenstein series, and $E^\star_\mathfrak{a} (z,s)$ are the completed Eisenstein series of $\Gamma_0(2)$. Using the Kronecker limit formul\ae\ \eqref{kronecker} and the duplication formul\ae
\begin{equation}
\vartheta_4 (\tau ) = \frac{\eta^2 (\tau /2)}{\eta (\tau)}\,, \qquad 
\vartheta_2 (\tau ) = \frac{\eta^2 (2 \tau )}{\eta (\tau)} \,,
\end{equation}
with $\vartheta_\alpha (\tau)$ the Jacobi theta constants, 
one may readily extract the residue and write
\begin{equation}
{\rm R.N.}\int_{\cF_2}\de\mu\,  \varGamma_{(2,2)} \left[ {\textstyle{0\atop 1}}\right] (U,T) = -\log \left( \frac{\pi\, e^{-\gamma}}{4}\,
U_2 \, T_2\,|\vartheta_2 (U) \, \vartheta_4 (T) |^4 \right)\,. \label{resultgamma2}
\end{equation}
This result agrees with the fact that this $\mathbb{Z}_2$ shift orbifold is equivalent to a (unshifted) two-dimensional Narain lattice with moduli $T/2$ and $2U$. Notice that, as expected, the result \eqref{resultgamma2} is not invariant under the full ${\rm O} (2,2;\mathbb{Z})$ $T$-duality group, but only under its subgroup $\Gamma_0(2)_U \times \Gamma^0(2)_T \ltimes \mathbb{Z}_2$, where $\Gamma^0(2)$ is the  congruence subgroup of level 2 of ${\rm SL} (2;\mathbb{Z})$
defined by $b=0\ {\rm mod}\ 2$, and $\mathbb{Z}_2$ maps $T\leftrightarrow -1/U$.

\section{One-loop BPS amplitudes from Niebur-Poincar\'e series \label{sec_np}}

We now turn to a different class of modular integrals of the form
\be
\label{defI2}
\cI_{(d,d+k)} ={\rm R.N.}\, \int_{\cF_N} \de\mu \,  \frac{1}{N}
\sum_{{g=1\atop (g,N)=1}}^{N-1} \varGamma_{(d,d+k)}\ar{0}{g} \, \varPhi \ar{0}{g} (\tau)\,,
\ee
of interest in heterotic-string compactifications.  Here, $\varGamma_{(d,d+k)} \ar{0}{g} $ is the Narain partition function \eqref{Narainlattice} associated to an even lattice of signature $(d,d+k)$, parametrised by the usual torus moduli $G_{ij}$ and $B_{ij}$ and by $k$ Wilson lines $Y_i^a$,  invariant under $\Gamma_0(N)\times O_{\vec \lambda} (d,d+k;\mathbb{Z})$.  The asymmetry in the signature of the lattice  implies that $\varPhi \ar{0}{g} (\tau )$ is a modular form of $\Gamma_0(N)$ with negative weight $w=-k/2$. 
Actually, for BPS-saturated amplitudes, that control the moduli dependence of gauge and gravitational threshold corrections in the heterotic string, the modular function $\varPhi \ar{0}{g} (\tau )$ is a weak (almost) holomorphic modular form related to the elliptic genus.  Holomorphy is a direct consequence of the fact that only half-BPS states, characterised by a non-excited right-moving vacuum, contribute to the amplitude, and this is the case for $T^4/\mathbb{Z}_N$ orbifolds with $N=2,3,4,6$. In these cases, the sum in \eqref{defI2} contains at most two terms with $g=1$ and $g=N-1$, and, moreover, $\varPhi \ar{0}{1} = \varPhi \ar{0}{N-1} \equiv \varPhi (\tau )$ since the two sectors are conjugate to each other\footnote{The same also holds for the Narain partition functions $\varGamma_{(d,d+k)} \ar{0}{g} $.}, thus yielding the integral
\be
\label{defI3}
\cI_{(d,d+k)} = {\rm R.N.}\, \int_{\cF_N} \de\mu \,\frac{1}{N} 
\sum_{{g=1\atop (g,N)=1}}^{N-1} \varGamma_{(d,d+k)}\ar{0}{g} \, \varPhi  (\tau)\,,
\ee
which can now be computed using the unfolding procedure outlined in Section \ref{genunfolding}.
To preserve manifest invariance under the T-duality group $O_{\vec\lambda} (d,d+k;\mathbb{Z})$, 
one should be able to represent the negative-weight modular form $\varPhi$ as an absolutely convergent Poincar\'e series. As in our previous 
work \cite{Angelantonj:2012gw}, this will be achieved by representing it as a linear combination of Niebur-Poincar\'e series. 

\subsection{Niebur-Poincar\'e series for $\Gamma_0(N)$}

Generalising the approach in \cite{Angelantonj:2012gw}, a convenient class of absolutely convergent Poincar\'e series of $\Gamma_0 (N)$ with negative modular weight $w$ is provided by 
 the Niebur-Poincar\'e series \cite{0288.10010,0543.10020,1004.11021} attached to the cusp $\mathfrak{a}$
\begin{equation}
\label{Fskw}
\cF_\mathfrak{a} (s,\kappa,w;\tau) = \tfrac12 \sum_{\gamma\in \varGamma_\mathfrak{a} \backslash \Gamma_0(N)} \,
\cM_{s,w}(-\kappa\tau_2)\, e^{-2\pi\I\kappa\tau_1}\, \Big\vert_w\, \sigma_\mathfrak{a}^{-1}\, \gamma \,.
\end{equation}
As in \cite{Angelantonj:2012gw}, $\cM_{s,w}$ is expressed in terms of the Whittaker $M$-function 
\begin{equation}
\cM_{s,w}(y) = |4\pi y|^{-\frac{w}{2}}\, M_{\frac{w}{2}\sgn(y), s-\frac12}
\left(4\pi |y| \right)\,,
\label{curlyM}
\end{equation}
so that the Poincar\'e series converges absolutely for ${\rm Re}\,(s)>1$. $\cF_\mathfrak{a} (s,\kappa,w)$ has a pole of order $\kappa$ in $q$ at the cusp  $\mathfrak{a}$, while being regular at all other cusps. All Niebur-Poincar\'e series are eigenmodes of the weight-$w$ Laplacian on $\mathbb{H}$,
\begin{equation}
\label{laplFskw}
\left[\Delta_w + \tfrac{1}{2}\, s(1-s) +\tfrac{1}{8}\, w(w+2)\right] \, \cF_\mathfrak{a} (s,\kappa,w) = 0 \,,
\end{equation}
as a consequence of the specific form of the seed function. Particularly interesting are the cases $s=\frac{w}{2}$ and $s=1-\frac{w}{2}$ for which $\cF_\mathfrak{a}$ becomes a harmonic Maass form. Notice that the latter choice will allow us to represent any weak holomorphic modular form of negative weight $w$, in terms of absolutely convergent Poincar\'e series\footnote{Special care is required for the case $w=0$, which can be defined by analytic continuation.}.

Other values of interest are those where $s=1-\tfrac{w}{2}+n$ with $n$  integer, since they are associated to weak almost holomorphic modular forms. These can be reached from $\cF_\mathfrak{a} (1-w/2,\kappa,w)$ via the action of the ladder operators
\be
\label{modularderiv}
D_w=\frac{\I}{\pi} \left( \partial_\tau-\frac{\I w}{2\tau_2} \right) \ ,\qquad
\bar D_w=-\I \pi\, \tau_2^2 \,\partial_{\bar\tau}\,,
\ee
which change the modular weight by units of $2$,
\begin{equation}
\label{DwF}
\begin{split}
D_w\cdot \cF_\mathfrak{a} (s,\kappa,w) &= 2\kappa\, (s+\tfrac{w}{2})\,   \cF_\mathfrak{a} (s,\kappa,w+2)\,,
\\
\bar D_w \cdot \cF_\mathfrak{a}  (s,\kappa,w) &= \frac{1}{8\kappa} (s-\tfrac{w}{2})\,   \cF_\mathfrak{a}  (s,\kappa,w-2)\,.
\end{split}
\end{equation}
Recall that the operator $\bar D_w$ maps a harmonic Maass form of weight $w$ to $\tau_2^{2-w}$ times (the complex conjugate of) a holomorphic cusp form of weight $2-w$, known as the shadow. Hence, if the shadow vanishes, the harmonic Maass form is in fact (weakly) holomorphic.  

For general $s$ with ${\rm Re}\,(s)>1$, the Fourier expansion of $\cF_{\mathfrak{a}}$ at the cusp $\mathfrak{b}$ is  given by \cite{1154.11015,BringmannOno2008}
\begin{equation}
\label{FskwF}
\cF_{\mathfrak{a}}(s,\kappa,w;\tau)\big\vert_{w}\sigma_{\mathfrak{b}}
=\delta_{\mathfrak{ab}}\, \cM_{s,w}(-\kappa \tau_2)\, e^{-2\pi\I\kappa\tau_1}
+ \sum_{m\in\IZ} \, \tilde\cF_{\mathfrak{ab}}(s,\kappa,w;\tau_2,m) \, e^{2\pi\I m\tau_1}\,,
\end{equation}
where, for zero frequency
\begin{equation}
\tilde\cF_{\mathfrak{ab}}(s,\kappa,w;\tau_2,0) =\frac{2^{2-w}\, \I^{-w}\, \pi^{1+s-\frac{w}{2}}\, \kappa^{s-\frac{w}{2}}\, \varGamma (2s-1) \cZ_{\mathfrak{ab}}(0,-\kappa;s)}{\varGamma(s-\frac{w}{2}) \, \varGamma(s+\frac{w}{2})}
\, \tau_2^{1-s-\frac{w}{2}}\,,
\label{Fzero}
\end{equation}
while for non-vanishing integer frequencies\footnote{Note that $\tilde\cF_{-\kappa<0}$  does {\it not} include the contribution from the first term in \eqref{FskwF}.}
\begin{equation}
\tilde\cF_{\mathfrak{ab}}(s,\kappa,w;\tau_2,m)  =\frac{4\pi\kappa\, \I^{-w}\, \varGamma(2s)}{ \varGamma ( s+\frac{w}{2} \, {\rm sgn}(m))}\, 
 \left|  \frac{m}{\kappa}\right|^{\frac{w}{2}} \, 
 \cZ_{\mathfrak{ab}}(m,-\kappa;s)\, \cW_{s,w}(m\tau_2)\,,
\label{FourierFBO}
\end{equation}
with $\cW_{s,w} (y)= |4\pi y|^{-\frac{w}{2}}\, W_{\frac{w}{2}{\rm sgn} (y), s-\frac{1}{2}} (4\pi |y|) $, expressed in terms of the Whittaker $W$-function.
As usual, $ \cZ_{\mathfrak{ab}}(m,n;s)$ is the Kloosterman-Selberg zeta function defined in \eqref{Zivgen}.
In particular, $\cF_{\mathfrak{a}}(s,\kappa,w)$ grows exponentially at the cusp 
$\mathfrak{a}$, but only as a power  $\tau_2^{1-s-w/2}$ at the other cusps.

For $w<0$, the special value $s=1-w/2$ lies inside the domain of absolute convergence, 
and the Fourier expansion \eqref{FskwF}
takes the expected form for a harmonic Maass form of weight $w$,
\begin{equation}
\begin{split}
\cF_{\mathfrak{a}}(1-\tfrac{w}{2},\kappa,w) \big|_w \sigma_\mathfrak{b}= & \delta_{\mathfrak{ab}}\, \left[ \varGamma (2-w) + (1-w)\, \varGamma (1-w;4\pi\kappa \tau_2 ) \right] q^{-\kappa}
\\
&+ \sum_{m\in\IZ} \, \tilde\cF_{\mathfrak{ab}}(1-\tfrac{w}{2},\kappa,w;\tau_2,m)  
\, e^{2\I \pi m \tau_1}\,, 
\end{split}
\end{equation}
where \begin{equation}
\label{FourierFBOh}
\begin{split}
\tilde\cF_{\mathfrak{ab}}(1-\tfrac{w}{2},\kappa,w;\tau_2,m>0)&= 4\pi\kappa\,  \I^{-w} \, \varGamma(2-w)\, 
\left(  \frac{m}{\kappa}\right)^{\frac{w}{2}}\, \cZ_{\mathfrak{ab}}(m,-\kappa;1-\tfrac{w}{2}) 
\, e^{-2\pi m\tau_2}\,,
\\
\tilde\cF_{\mathfrak{ab}}(1-\tfrac{w}{2},\kappa,w;\tau_2,m<0)&=4\pi\kappa\,  \I^{-w} \, (1-w)\left(  \frac{-m}{\kappa}\right)^{\frac{w}{2}} 
\cZ_{\mathfrak{ab}}(m,-\kappa;1-\tfrac{w}{2})
\\
&\quad \times 
\varGamma(1-w,-4\pi m\tau_2)\, e^{-2\pi m\tau_2}\,,
\\
\tilde\cF_{\mathfrak{ab}}(1-\tfrac{w}{2},\kappa,w;\tau_2,m=0)&=\frac{4\pi^2\,\kappa}{(2\pi\I\kappa)^w} \,
 \cZ_{\mathfrak{ab}}(0,\kappa;1-\tfrac{w}{2}) \,.
 \end{split}
 \ee
Applying the lowering operator $\bar D_w$, one finds that the shadow of this harmonic Maass form is the holomorphic Poincar\'e series $P(-\kappa,2-w)$ of $\Gamma_0 (N) $ with weight $2-w$. If the space of cusp forms $\cS_w(N)$ is trivial, this Poincar\'e series must vanish and therefore $\cF_{\mathfrak{a}}(1-\frac{w}{2},\kappa,w)$ is  an ordinary weak holomorphic modular form of weight $w$.  More generally, it is straightforward to prove that, 
the shadow of a linear combination 
\be
\label{Bruinierexp}
 \sum_{\mathfrak{a}}
\sum_{-\kappa_{\mathfrak{a}}\leq \ell <0} 
c_{\mathfrak{a},\ell} \, \cF_{\mathfrak{a}}(1-\tfrac{w}{2},\ell ,w) 
\ee
vanishes whenever $\sum_{-\kappa_{\mathfrak{a}}\leq \ell<0}
c_{\mathfrak{a},\ell} \ \, q^{\ell }$  corresponds to the principal part of a weak holomorphic modular form $\varPhi_w$ of $\Gamma_0 (N)$
at the cusp $\mathfrak{a}$. In this case, the linear combination \eqref{Bruinierexp} in fact coincides with $\varPhi_w$ itself. As a result, one can use this property to express a generic weak holomorphic modular form $\varPhi_w$ of weight $w$ with principal part
\be
\varPhi_w ^- = \sum_{-\kappa_\mathfrak{a} \le \ell <0} c_{\mathfrak{a},\ell} \, q^{\ell }
\ee
at the cusp $\mathfrak{a}$, as the linear combination
\be
\varPhi_w = \frac{1}{\varGamma (2-w)}\sum_\mathfrak{a} \sum_{-\kappa_\mathfrak{a} \le \ell <0} c_{\mathfrak{a},\ell}\, \cF_\mathfrak{a} (1-\tfrac{w}{2},\ell,w )\,,
\label{generalehxpansion}
\ee
of Niebur-Poincar\'e series at the various cusps.
 For $w=0$, the value $s=1-w/2=1$ belongs to the boundary of the convergence domain, and the value of the Kloosterman-Selberg zeta function $\cZ_{\mathfrak{ab}}(m,n;s)$ must be defined by analytic continuation.  The previous equality \eqref{generalehxpansion} holds up to an additive constant.
 
For $s=1-\frac{w}{2} +n$, with $n$ a positive integer, the seed of the Niebur-Poincar\'e series reads
 \be
 \begin{split}
 \cM_{1-\frac{w}{2}+n,w} (-\kappa \tau_2 ) \, e^{-2\pi \I \kappa \tau_1 } &= \varGamma (2n+2-w) \, (4\pi \kappa \tau_2)^{-n} 
 \\
 &\qquad \times \left[ q^{-\kappa}\, L^{(-1-2n+w)}_n (-4\pi\kappa \tau_2 ) - \bar q ^{\kappa}\, L^{(-1-2n+w)}_{n-w} (4\pi\kappa \tau_2)\right] \,,
 \end{split}
 \ee
where $L^{(\alpha)}_m (x)$ are the generalised Laguerre polynomials, and thus $\cF_\mathfrak{a} (1-\frac{w}{2}+n,\kappa,w)$ can be used to represent weak almost holomorphic modular forms. In fact, if a generic weak almost holomorphic modular form $\varPsi_w$ of weight $w$, containing at most $n$ powers of $\hat E_2$ factors, has principal part\footnote{By abuse of language, we consider here the bare $\tau_2$-factors as independent variables, and we expand the weak  almost holomorphic modular forms in powers of $q$ with $\tau_2$-dependent coefficients.} 
\be
\varPsi_w^{-} = \sum_{-\kappa_\mathfrak{a} \le m < 0}\sum_{\ell=0}^n \frac{c_{\mathfrak{a},\ell} (m)\, \tau_2^{\ell -n}}{q^m}
\ee
near the cusp $\mathfrak{a}$, then it can be uniquely  decomposed as
\be
\varPsi_w = \sum_\mathfrak{a} \sum_{-\kappa_\mathfrak{a} \le m <0}\sum_{p=0}^n \, d_{\mathfrak{a},p} (m )\, \cF_\mathfrak{a} (1-\tfrac{w}{2}+p,m,w)
\label{generalnhw}
\ee
plus, eventually, a constant in the case $w=0$. The coefficients $d_{\mathfrak{a},p} (m)$ are then determined recursively in terms of $c_{\mathfrak{a},\ell}(m)$, by comparing the principal parts of both sides of Eq. \eqref{generalnhw},
\be
\begin{split}
d_{\mathfrak{a},n} (m) &= \frac{c_{\mathfrak{a},0} (m)}{A_{n,0}}\,,
\\
d_{ \mathfrak{a}, n-p} (m  ) &= \frac{c_{\mathfrak{a}, p} (m) - \sum_{\ell =n-p+1}^{n} A_{\ell ,p}\, d_{\mathfrak{a},\ell} (m)}{A_{n-p,p}}\,,
\end{split}\label{nhsolution}
\ee
where
\be
A_{\ell, p} =\frac{\varGamma(2\ell +2-w) \,  \varGamma (\ell -w+n-p +1)}{\varGamma (\ell+1-w)\, \varGamma (n-p +1) \, \varGamma (p+\ell -n +1)}
\, (-4\pi m)^{p-n}\,.
\ee

We close this subsection by identifing the Niebur-Poincar\'e series 
$\cF_{\mathfrak{a}}(s,\kappa,w)$ at 
$s=1-w/2+n$ in terms of ordinary (almost) holomorphic modular forms
defined in Appendix \ref{sec_comp}.  In order to avoid confusion, we shall introduce a new label to display the level $N$ of the congruence subgroup $\Gamma_0(N)$, {\em i.e.} $\cF_{\mathfrak{a}}^{(N)}$ will denote the Niebur-Poincar\'e series of $\Gamma_0(N)$. For $\Gamma_0(2)$, we find for example
\be
\begin{split}
&\cF^{(2)}_{\infty}(1,1,0)=J_2-8\,,\quad\cF^{(2)}_{\infty}(1,2,0)=J_2^2-544\,,\quad \cF^{(2)}_{\infty}(1,3,0)=J_2^3-828 J_2+6112\,,
\\
&\cF^{(2)}_{\infty}(2,1,-2)=2\frac{E_6-2 X_2 E_4}{\Delta_8^{(2)}}\,,\quad
\cF^{(2)}_{\infty}(2,2,-2)=\frac{X_2(7E_4-31 X_2^2)(E_4-4X_2^2)^2}{36 (\Delta_8^{(2)})^2}\,,
\\
&\cF^{(2)}_{\infty}(3,1,-4)=40\frac{4X_2^2-E_4}{\Delta_8^{(2)}}\, ,\quad
\cF^{(2)}_{\infty}(3,2,-4)=-20\frac{(E_4-7X_2^2)(E_4-4X_2^2)^2}{9 (\Delta_8^{(2)})^2}\,,
\\
&\cF^{(2)}_{\infty}(2,1,0)=\frac12 D \cdot \left [ \frac{E_6-2 X_2 E_4}{3\Delta_8^{(2)}} \right] = 
\frac{(E_4-4 X_2^2)(E_4-6X_2^2+\hat E_2 X_2)}{18 \Delta_8^{(2)}}\,.
\end{split}
\ee
Note that Niebur-Poincar\'e series with weight $w<-4$ under $\Gamma_0(2)$
are genuine harmonic Maass forms. Niebur-Poincar\'e series associated to the cusp $0$ can be obtained using Eq. \eqref{conjugaterelation} and the modular properties of the almost holomorphic modular forms. 

For $\Gamma_0(3)$, we find instead
\be
\begin{split}
&\cF^{(3)}_{\infty}(1,1,0)=J_3-3\, ,\quad
\cF^{(3)}_{\infty}(1,2,0) = J_3^2-117\, ,\quad \cF^{(3)}_{\infty}(1,3,0)=J_3^3-162 J_3+243\,,
\\
&\cF^{(3)}_{\infty}(2,1,-2)=3\frac{9 X_3^2- 4E_4}{16\Delta_6^{(3)}}\, ,\quad
\cF^{(3)}_{\infty}(2,2,-2)=\frac{256 E_4 E_6-1053 X_3^5+1368 E_4 X_3^3-784 E_4^2 X_3}{4096(\Delta_6^{(3)})^2}\,.
\end{split}
\ee
Similarly, Niebur-Poincar\'e series with weight $w<-2$ under $\Gamma_0(3)$
are genuine harmonic Maass forms. Also in this case, Niebur-Poincar\'e series associated to the cusp $0$ can be obtained using Eq. \eqref{conjugaterelation} and the modular properties of the holomorphic modular forms. 

\subsection{BPS-state sums}

Since any weak almost  holomorphic modular form of negative weight can be represented 
as a linear combination of Niebur-Poincar\'e series $\cF_\mathfrak{a} (s,\kappa,w)$,  it suffices to consider the basic integral
\begin{equation}
\label{mainint}
{\mathcal I}_{d+k,d}  (s,\kappa; \mathfrak{a} )
\equiv {\rm R.N.} \, \int_{\cF_N} \de\mu\, \frac{1}{N} 
\sum_{{g=1\atop (g,N)=1}}^{N-1} \varGamma_{(d,d+k)}\ar{0}{g} \,  \cF_\mathfrak{a} (s, \kappa , -\tfrac{k}{2})\,,
\end{equation}
where the modular weight  $w=-k/2$ is determined by the signature of the Narain lattice, and the definition of the renormalised integral follows from Eq. \eqref{generalRNint}.

Notice that, as in \cite{Angelantonj:2012gw}, the proper definition of the renormalised integral requires some care since for special values of $s$ the second integral in the {\em r.h.s.} of Eq. \eqref{generalresult} can develop a simple pole. For instance, the contribution of the cusp $\infty$ reads
\begin{equation}
f_\mathfrak{a\infty} (s) \, \frac{\cT^{\frac{2d+k}{4}-s}}{s-\frac{2d+k}{4}}\,,
\end{equation}
where $f_\mathfrak{a\infty} (s)$ is the $\tau_2$ independent part in \eqref{Fzero}. If the lattice-shift $\vec\lambda \not=0$ this is actually the only divergent contribution since the lattice is exponentially suppressed for $\mathfrak{b}\not=\infty$. If however the shift is trivial $\vec \lambda =0$, then similar contributions may originate also from the other cusps.

Following the unfolding procedure outlined in Section \ref{genunfolding} one thus finds  
\begin{equation}
\begin{split}
{\mathcal I}_{d+k,d}  (s,\kappa; \mathfrak{a} ) &= \frac{1}{N} \sum_{{g=1\atop (g,N)=1}}^{N-1} {\mathcal I}_{d+k,d}  (s,\kappa; \mathfrak{a} ) \ar{0}{g}
\\
&\equiv \frac{1}{N} \sum_{{g=1\atop (g,N)=1}}^{N-1} 
\int_\cS \de\mu\,\varGamma_{(d,d+k)}\ar{0}{g} (\sigma_\mathfrak{a}\tau) \, j^{-k/2}_{\sigma_\mathfrak{a}} (\tau)\, 
\cM_{s,-\frac{k}{2}} (-\kappa\tau_2 ) \,e^{-2\I\pi\kappa\tau_1}\,.
\end{split} \label{intNPlattice}
\end{equation}
One can further simplify the result by noting that 
\begin{equation}
 j^{-k/2}_{\sigma_\mathfrak{a}} (\tau)\, \varGamma_{(d,d+k)}\ar{0}{g} (\sigma_\mathfrak{a}\tau) = m_\mathfrak{a}^{k/4}\, 
\varGamma _{(d,d+k)}\ar{v \, g}{u\,g} (m_\mathfrak{a} \tau )\,,
\end{equation}
where $m_\mathfrak{a}$ is the width of the cusp $\mathfrak{a}$ associated to the rational number $u/v$ ($(u,v)=1$), and we have made use of the modular transformation property \eqref{modulartransformation} with $\gamma = \tau_\mathfrak{a} = {\scriptstyle\left( \begin{array}{cc} u & * \\ v & * \end{array}\right)}$ applied to the Narain lattice with characteristics. Using the definition \eqref{Narainlattice}, and the relations
\begin{equation}
 p_{\rm L}^2 (h)  + p_{\rm R}^2 (h)  = 2\, \cM^2 (h) 
\,,\qquad
p_{\rm L}^2 (h)- p_{\rm R}^2 (h) = 2 \, (\vec p + h \vec \lambda )^2 \,,
\end{equation}
that can be seen as a definition of the left-moving and right-moving momenta, one finds
\begin{equation}
\begin{split}
{\mathcal I}_{d+k,d}  (s,\kappa; \mathfrak{a} ) \ar{0}{g} &= m_\mathfrak{a}^{\frac{2d+k}{4}} \sum_{\vec p} e^{2\I\pi ug \vec\lambda\cdot\vec p}\int_0^\infty \de\tau_2\, \tau_2^{d/2-2} e^{-\pi m_\mathfrak{a}\tau_2 \cM^2 (v g)}\, \cM_{s,-\frac{k}{2}} (-\kappa \tau_2 ) 
\\
&\qquad\times \int_0^1 \de\tau_1\, e^{2\I\pi \tau_1 ( \frac{m_\mathfrak{a}}{2} (\vec p+ v g \vec \lambda )^2 - \kappa )}
\\
&= m_\mathfrak{a}^{\frac{2d+k}{4}}\, (4\pi\kappa )^{1-d/2} \, \varGamma \left( s+\tfrac{2d+k}{4}-1\right) 
\\
&\times\sum_{\rm BPS} e^{2\I\pi ug \vec\lambda\cdot\vec p}\, {}_2 F_1 \left( s-\tfrac{k}{4}, s+ \tfrac{2d+k}{4}-1 ;2s;\tfrac{4\kappa}{m_\mathfrak{a} p_{\rm L}^2 (vg)}\right) \, \left(\frac{m_\mathfrak{a} p_{\rm L}^2 (vg)}{4\kappa}\right)^{1-s-\tfrac{2d+k}{4}}\,,
\end{split}
\end{equation}
where now 
\begin{equation}
\sum_{{\rm BPS}(h)} \equiv \sum_{p_{\rm L}(h) ,p_{\rm R}(h)} \delta (4\kappa-m_\mathfrak{a} [p_{\rm L}^2 (h)- p_{\rm R}^2 (h)]) =
\sum_{\vec p \in \mathbb{Z}^{2d+k}} \delta (4\kappa- 2 m_\mathfrak{a} (\vec p +h\vec\lambda)^2) \,.
\end{equation}
Putting things together, the integral \eqref{intNPlattice} evaluates to
\begin{equation}
\begin{split}
{\mathcal I}_{d+k,d}  (s,\kappa; \mathfrak{a} ) &= \frac{1}{N} \sum_{{g=1\atop (g,N)=1}}^{N-1}
m_\mathfrak{a}^{\frac{2d+k}{4}}\, (4\pi\kappa )^{1-d/2} \, \varGamma \left( s+\tfrac{2d+k}{4}-1\right) 
\\
&\times\sum_{{\rm BPS} (vg )} e^{2\I\pi ug \vec\lambda\cdot\vec p}\, {}_2 F{}_1 \left( s+ \tfrac{2d+k}{4}-1,s-\tfrac{k}{4};2s;\tfrac{4\kappa}{m_\mathfrak{a} p_{\rm L}^2 (vg)}\right) \, \left(\frac{m_\mathfrak{a} p_{\rm L}^2 (vg)}{4\kappa}\right)^{1-s-\tfrac{2d+k}{4}}\,.
\end{split}\label{finalintegral00}
\end{equation}
The BPS-sum is absolutely convergent for ${\rm Re}\, (s)$ sufficiently large, and admits a meromorphic continuation to ${\rm Re}\, (s) >1$ with a simple pole at $s=\frac{k+2d}{4}$. In this case, the renormalised integral is given by the constant term in the Laurent expansion of \eqref{finalintegral00} at this point.

Notice that this expression coincides with the one in \cite{Angelantonj:2012gw} obtained in the case of integrals (and integrands) associated the full modular group, modulo the twist-dependent phase the appearance of the width $m_\mathfrak{a}$ and an explicit dependence of the left-moving and right-moving momenta on the twist. Therefore, for integer values of $s$,  the case of main interest in this paper, the result simplifies in terms of elementary functions, similarly to Eqs. (3.28--32) in \cite{Angelantonj:2012gw}. Moreover, the modular integral can be straightforwardly evaluated even in the presence of non-trivial ($\Gamma_0 (N)$ invariant) insertions of left-moving and right-moving momenta, similarly to \cite{Angelantonj:2012gw}.

\section{Examples \label{sec_examples}}

We close this paper with some examples drawn from heterotic strings compactified on ${\rm K}3\times T^2$ freely-acting orbifolds. In these cases, the original $\cN =4$ supersymmetry is spontaneously broken down to $\cN =2$, and the scale of (partial) supersymmetry breaking is set by the size of the compact dimensions orthogonal to K3. We shall also discuss a notable example of type II string thermodynamics where our methods can be used to compute the free energy.

\subsection{$\cN=2$ heterotic string vacua}

Threshold corrections to gauge and  gravitational couplings in heterotic string vacua with $\cN=2$ supersymmetry in 4 dimensions are given by the one-loop integrals  \cite{Antoniadis:1992rq,Dixon:1990pc,Harvey:1995fq,Harvey:1996gc}
\be
\label{thresh}
\begin{split}
\Delta_{G} =& -\I \int_{\cF} \de\mu
\frac{\tau_2 }{2\eta^2} {\rm Tr} \left\{ J_0 \, e^{\I \pi J_0} \, q^{L_0-\tfrac{c}{24}}\, 
\bar q^{\bar L_0-\tfrac{\bar c}{24}}\, \left( Q^2 - \frac{1}{4\pi\tau_2} \right)\right\} 
\,,
\\
\Delta_{\rm grav} =&  -\I \int_{\cF} \de\mu
\frac{\tau_2 }{2\eta^2} {\rm Tr} \left\{ J_0 \, e^{\I \pi J_0} \, q^{L_0-\tfrac{c}{24}}\,
\bar q^{\bar L_0-\tfrac{\bar c}{24}}\right\}  \frac{\hat E_2}{12}
\,,
\end{split}
\ee
where the traces run over the internal $(c,\bar{c})=(9,22)$ superconformal field theory, with the right-movers in the Ramond ground state and $J_0$ being the total ${\rm U}(1)$ generator of the $c=9$ superconformal algebra. The quantity appearing in braces in the second line 
is the modified elliptic genus $Z$, a modular form of weight $(-2,0)$. $Q$ denotes one of the Cartan generators in the gauge group $G$. These integrals are clearly ill-defined since they suffer from IR divergences. Although the traditional way to cope with these divergences is to explicitly subtract  the contribution $b_{{\rm grav}/G} \, \tau_2$ of the massless states, with $b_{\rm grav}, b_{G}$  being the coefficients of the corresponding one-loop beta functions, we shall employ here a different renormalisation prescription associated to the definition of the renormalised integral \eqref{generalRNint}. 
 
We focus on heterotic compactifications on $(T^2\times T^2\times T^2)/\IZ_N$, where $\IZ_N$, $N=2,3,4,6$,
acts as a rotation $(z_1,z_2,z_3)\mapsto (e^{2\pi\I/N} z_1,e^{-2\pi\I/N} z_2,z_3)$ on the first two complex coordinates, times an order $N$  translation on the remaining $T^2$ and on the internal ${\rm E}_8\times {\rm E}_8$ lattice, parametrised by constant integer vectors $\gamma^I, \gamma'^I$, 
$I=1\dots 8$, satisfying the level matching condition $\gamma^2+\gamma'^2-2 = 0 \,({\rm mod}\ 2N)$. 
In the absence of translation along $T^2$, this is simply the heterotic string compactified
on ${\rm K3}\times T^2$. For such freely acting heterotic orbifolds,
and for vanishing Wilson lines, the modified elliptic genus is given by a  sum 
 \be
Z= \frac{\tau_2}{\eta^2}\,{\rm Tr} \left\{ J_0 \, e^{\I \pi J_0} \, q^{L_0-\tfrac{c}{24}}\, 
\bar q^{\bar L_0-\tfrac{\bar c}{24}}\right\} = \frac{1}{N}\sum_{h,g\in\mathbb{Z}_N} Z\ar{h}{g}\ ,\quad 
\ee
where \footnote{For clarity, we will explicitly display between parentheses the level $N$ of the congruence subgroup $\Gamma_0(N)$ of quantities in question.}
\be
Z\ar{h}{g}= \frac{\I}{\eta^{20}(\tau)}\,
Z^{\rm K3}\ar{h}{g}\, Z^{{\rm E}_8\times {\rm E}_8}\ar{h}{g}\, \varGamma_{(2,2)}\ar{h}{g}
\ee
is a product of the holomorphic orbifold blocks of K3,
\be
Z^{\rm K3}\ar{h}{g} = k\ar{h}{g}\,\frac{\eta^2}{\theta\bigr[^{~\frac{1}{2}+\frac{h}{N}~}_{~\frac{1}{2}+\frac{g}{N}~}\bigr]\,\theta\bigr[^{~\frac{1}{2}-\frac{h}{N}~}_{~\frac{1}{2}-\frac{g}{N}~}\bigr]} \ ,
\ee
the holomorphic orbifold blocks of the ${\rm E}_8\times {\rm E}_8$ lattice
\be
Z^{{\rm E}_8\times {\rm E}_8}\ar{h}{g}=e^{-2\pi\I hg \sum_{I=1}^8 (\gamma_I^2+\gamma'^2_I)/  N^2}\, 
Z^{{\rm E}_8}\ar{h}{g}\, Z^{{\rm E}'_8}\ar{h}{g}\,,
\ee
where 
\be
Z^{{\rm E}_8}\ar{h}{g}=\tfrac{1}{2} \sum_{k,\ell=0}^1 e^{-\I\pi h\ell  (\sum_{I=1}^8 \gamma^I)/N}\,
\prod_{I=1}^8 \theta\left[ \begin{array}{c} \frac{k}{2}+\frac{h}{N} \gamma^I \\ \frac{\ell}{2}+\frac{g}{N} \gamma^I
\end{array}\right] 
\ee
and, finally, the shifted Narain lattice partition function $\varGamma_{(2,2)}\ar{h}{g}$, defined in \eqref{Narainlattice}. The  $k\ar{h}{g}$'s are numerical constants determined by $k\ar{0}{g}=16\,\sin^4(\pi g/N)$ and by modular invariance\footnote{Their explicit expressions can be found, for instance, in \cite{Henningson:1996jz,Stieberger:1998yi}}. In the absence of a shift along the third torus, the lattice is independent of $h,g$ and one
finds, for all models, a unique answer determined by modular invariance 
\be
Z= -2\I \frac{E_4 E_6}{\eta^{24}} \,\varGamma_{(2,2)}\ .
\ee
The integrals \eqref{thresh} can then be computed using the usual unfolding of the fundamental domain
of ${\rm SL}(2;\IZ)$. For freely acting orbifolds, with a non-trivial order-$N$ shift along the third $T^2$, 
the integrals \eqref{thresh} are best computed by unfolding the fundamental domain of 
$\Gamma_1(N)$, or possibly some larger level $N$ subgroup of ${\rm SL}(2;\IZ)$ (see \cite{Mayr:1993mq} for an early application of this technique). 
For the standard embedding, corresponding to 
\be
\gamma=(1,-1,0,0,0,0,0,0), \quad \gamma'=(0,0,0,0,0,0,0,0)\ ,
\ee
 one finds
 \be
 Z\ar{0}{1} = -2\I\, \frac{E_4 \varPhi_6}{\varDelta} \,\varGamma_{(2,2)}\ar{0}{1} \,,
 \ee
where $ \varPhi_6$ are modular forms of weight $6$ under $\Gamma_0(N)$, defined in terms of Jacobi theta functions by
\begin{equation}
	\varPhi_6 \equiv -4\,\eta^6\,\sin^4\tfrac{\pi}{N} \sum\limits_{k,\ell=0}^{1}\frac{ \theta[^{k/2}_{\ell/2}]^6 \,\theta[{k/2 \atop \ell/2+1/N}]\,\theta[{k/2 \atop \ell/2-1/N}]}{\theta[{1/2 \atop 1/2+1/N}]\, \theta[{1/2 \atop 1/2-1/N}]} ~,
\end{equation}
and, explicitly expressed  in terms of modular forms of $\Gamma_0(N)$, introduced in Appendix \ref{sec_comp}, as
\begin{equation}
\label{defphi62}
\begin{split}
\varPhi^{(2)}_6 =&\tfrac23 ( E_6 - 2 X_2 E_4) \,,
\\
\varPhi^{(3)}_6 =& \tfrac38 ( E_6 - \frac32 X_3 E_4) \,,
\\
\varPhi^{(4)}_6 =& \tfrac16 ( E_6 - (X_4 + 2 X_2(2\tau)) E_4)  \,,
\\
\varPhi^{(6)}_6 =& \tfrac{1}{24} ( E_6 - ( \tfrac92 X_2(3\tau) + 2 X_3(2\tau) + \tfrac12 X_6) E_4 ) \,.
\end{split}
\end{equation}
We have added here an index to keep track of the  level $N$ of the associated Hecke congruence subgroup $\Gamma_0 (N)$.
Therefore, for $N=2,3$ the gravitational threshold corrections are given by
\be
\Delta_{\rm grav} = -\frac{N-1}{12N} \, {\rm R.N.}\,
 \int_{\cF_N}\, \de\mu \, \frac{\hat E_2 \, E_4\, \varPhi_6^{(N)}}{\varDelta}\, \varGamma_{(2,2)}^{(N)}\ar{0}{1}\, ,
\ee
For $N=4$, one similarly gets the linear combination
\be
\Delta_{\rm grav} = 
 -\frac{1}{6} \, {\rm R.N.}\, \int_{\cF_4}\, \de\mu \, \frac{\hat E_2 \, E_4\, \varPhi^{(4)}_6}{\varDelta}\, \varGamma_{(2,2)}^{(4)}\ar{0}{1}\ 
 -\frac{1}{48}  \, {\rm R.N.}\,
 \int_{\cF_2}\, \de\mu \, \frac{\hat E_2 \, E_4\, \varPhi^{(2)}_6}{\varDelta}\, \varGamma^{(2)}_{(2,2)}\ar{0}{1}\ ,
\ee 
whereas for $N=6$ one gets
\be
\begin{split}
\Delta_{\rm grav} =& 
 -\frac{1}{36} \, {\rm R.N.}\,\int_{\cF_6}\, \de\mu \, \frac{\hat E_2 \, E_4\, \varPhi^{(6)}_6}{\varDelta}\, \varGamma_{(2,2)}^{(6)}\ar{0}{1}
-\frac{1}{36} \, {\rm R.N.}\,
 \int_{\cF_3}\, \de\mu \, \frac{\hat E_2 \, E_4\, \varPhi^{(3)}_6}{\varDelta}\, \varGamma_{(2,2)}^{(3)}\ar{0}{1}
 \\
 & 
 -\frac{1}{72} \, {\rm R.N.}\,
 \int_{\cF_2}\, \de\mu \, \frac{\hat E_2 \, E_4\, \varPhi^{(2)}_6}{\varDelta}\, \varGamma_{(2,2)}^{(2)}\ar{0}{1}\,.
 \end{split}
\ee 
The thresholds for the ${\rm E}_8$ and ${\rm E}_7$ gauge groups for $N=2,3$ are similarly given by
\be
\Delta_{{\rm E}_8} = -\frac{N-1}{12N} \, {\rm R.N.}\,
\int_{\cF_N}\, \de\mu \, \frac{(\hat E_2 \, E_4 - E_6)\, \varPhi^{(N)}_6}{\varDelta}\, \varGamma_{(2,2)}^{(N)}\ar{0}{1}
\ee
and
\be
\Delta_{{\rm E}_7} =-\frac{N-1}{12N} \, {\rm R.N.}\,
\int_{\cF_N}\, \de\mu \, \left[ \frac{(\hat E_2 \, E_4 - E_6 ) \, \varPhi^{(N)}_6}{\varDelta} - b_N \right]
\, \varGamma_{(2,2)}^{(N)}\ar{0}{1}\,,
\ee
with $b_2=1152$ and  $b_3=648$, while the thresholds for $N=4,6$ involve similar linear combinations
with $b_4=288$ and $b_6=72$.
It is worth noting that the modular form $\varPhi^{(N)}_6$ drops out from the difference 
$\Delta_{{\rm E}_8} -\Delta_{{\rm E}_7}$, e.g. for $N=2,3$
\be
\Delta_{{\rm E}_8} -\Delta_{{\rm E}_7}= 
-\frac{N-1}{12N}\, b_N  \, {\rm R.N.}\,
\int_{\cF_N}\, \de\mu \, \varGamma_{(2,2)}^{(N)}\ar{0}{1}\, .
\ee
In all these integrals, the integrand can be 
 represented as a linear combination of Niebur-Poincar\'e series. Let us see in some detail how it works in the $N=2$ case.
Using the definitions of the holomorphic Eisenstein series and of the $X_2$ modular form of $\Gamma_0 (2)$, given in Appendix B, one immediately finds the following behaviour near the cusp $\infty$
\be
\frac{E_6 \varPhi^{(2)}_6}{\varDelta} \Big| \sigma_\infty = \frac{2}{q} - 944 + \cO (q)\,. \label{expansion2infty}
\ee
To find the principal part of the Laurent expansion at the cusp $0$, we have to consider
\be
\begin{split}
\frac{E_6 \varPhi^{(2)}_6}{\varDelta} \Big| \sigma_0 &= \frac{2}{3} \frac{E_6 (2\tau ) \left[ E_6 (2\tau ) + X_2\, E_4 (2\tau)\right]}{\varDelta (2\tau )}
\\
&= -\frac{16}{q} -512 +\cO (q)\,.
\end{split}
\label{expansion20}
\ee
As a result, Eq. \eqref{generalehxpansion}, together with Eqs. \eqref{expansion2infty} and \eqref{expansion20}, yields
\be
\label{E62PhiDel}
\frac{E_6 \varPhi^{(2)}_6}{\varDelta} 
=  2\,\cF^{(2)}_{\infty}(1,1,0)-16\,\cF^{(2)}_{0}(1,1,0)-672 \,.
\ee
Similarly, 
\be
\frac{\hat E_2 E_4 \varPhi^{(2)}_6}{\varDelta}\Big| \, \sigma_\infty = \frac{2}{q} \left(1-\frac{3}{\pi\,\tau_2}\right) + 544 \left(1- \frac{3}{\pi \,\tau_2}\right) -48 +\cO(q)\,,
\ee
and
\be
\begin{split}
\frac{\hat E_2 E_4 \varPhi^{(2)}_6}{\varDelta}\Big| \, \sigma_0 &= \frac{2}{3} \frac{\hat E_2 (2\tau) \, E_4 (2\tau ) \left[ E_6 (2\tau ) + X_2\, E_4 (2\tau)\right]}{\varDelta (2\tau )}
\\
&= -\frac{16}{q} \left(1 -\frac{3}{\pi\, \tau_2}\right) -512 \left(1 -\frac{3}{\pi\,\tau_2}\right)+ \cO(q)\,,
\end{split}
\ee
so that Eqs. \eqref{generalnhw} and \eqref{nhsolution} yield
\be
\frac{\hat E_2 E_4 \varPhi^{(2)}_6}{\varDelta} = 
 -8 \,\cF^{(2)}_{0}(2,1,0)+2\,\cF^{(2)}_{\infty}(2,1,0) +32\,\cF^{(2)}_{0}(1,1,0)-10\,\cF^{(2)}_{\infty}(1,1,0)-96\,.
\ee
Following a similar procedure, one finds
\be
\label{E63PhiDel}
\begin{split} 
\frac{E_6 \varPhi^{(3)}_6}{\varDelta} 
= & \tfrac32 \left( \cF^{(3)}_{\infty}(1,1,0) - 9\, \cF^{(3)}_{0}(1,1,0)-3\, \cF^{(3)}_{0}(1,2,0)-252\right) \,,
\\
\frac{\hat E_2 E_4 \varPhi^{(3)}_6}{\varDelta} = & \tfrac38
\left( 4\,\cF^{(3)}_{\infty}(2,1,0)-8\,\cF^{(3)}_{0}(2,2,0)-12\,\cF^{(3)}_{0}(2,1,0) 
 -20\,\cF^{(3)}_{\infty}(1,1,0)\right.
 \\
 & \left. +36\,\cF^{(3)}_{0}(1,1,0) +36\, \cF^{(3)}_{0}(1,2,0)-144 \right)\,,
 \end{split}
 \ee
for $N=3$, 
\be
\label{E64PhiDel}
\begin{split} 
\frac{E_6 \varPhi^{(4)}_6}{\varDelta}= &
 \cF^{(4)}_{\infty}(1,1,0)-8\,  \cF^{(4)}_{0}(1,1,0)-4 \,\cF^{(4)}_{0}(1,2,0)
 -2 \,\cF^{(4)}_{0}(1,3,0)-168 \,,
\\
\frac{\hat E_2 E_4 \varPhi^{(4)}_6}{\varDelta}= &
-2\, \cF^{(4)}_{\infty}(1,1,0)+  \cF^{(4)}_{\infty}(2,1,0)
+70\, \cF^{(4)}_{0}(1,1,0)-2\, \cF^{(4)}_{0}(2,1,0) \\
&+23 \,\cF^{(4)}_{0}(1,2,0)-2 \,\cF^{(4)}_{0}(2,2,0)
+\tfrac{17}{2} \,\cF^{(4)}_{0}(1,3,0) - \tfrac32 \,\cF^{(4)}_{0}(2,3,0)
+336\,,
 \end{split}
\ee
for $N=4$, and finally
\be
\label{E66PhiDel}
\begin{split}
\frac{E_6 \varPhi^{(6)}_6}{\varDelta}= &\tfrac{1}{2} \left(
\cF^{(6)_{\infty}(1,1,0)}
+ \cF^{(6)}_{1/3}(1,1,0)
- \cF^{(6)}_{1/2}(1,1,0)
+ \cF^{(6)}_{1/2}(1,2,0) 
-6  \,\cF^{(6)}_{0}(1,1,0) \right.
\\
& \left. -4\,  \cF^{(6)}_{0}(1,2,0)
-3 \, \cF^{(6)}_{0}(1,3,0)
-2\, \cF^{(6)}_{0}(1,4,0)
- \cF^{(6)}_{0}(1,5,0) -84 \right)\,,
\\
\frac{\hat E_2 E_4 \varPhi^{(6)}_6}{\varDelta}= & 
\tfrac{1}{12}\,\left(
\mathcal{F}^{(6)}_\infty(2,1,0) 
- 30 \,\mathcal{F}^{(6)}_\infty(1,1,0) 
-5\,\mathcal{F}^{(6)}_0(2,5,0)
- 8\,\mathcal{F}^{(6)}_0(2,4,0)
-9\,\mathcal{F}^{(6)}_0(2,3,0) \right.
\\
& -8\,\mathcal{F}^{(6)}_0(2,2,0) 
-6\,\mathcal{F}^{(6)}_0(2,1,0)
+24\,\mathcal{F}^{(6)}_0(1,5,0)
+36\,\mathcal{F}^{(6)}_0(1,4,0)
+36\,\mathcal{F}^{(6)}_0(1,3,0)
\\
& +24\,\mathcal{F}^{(6)}_0(1,2,0)  
+ 4\,\mathcal{F}^{(6)}_{1/2}(2,2,0)
-2\,\mathcal{F}^{(6)}_{1/2}(2,1,0)
- 18\,\mathcal{F}^{(6)}_{1/2}(1,2,0)
 \\
&\left. +6\,\mathcal{F}^{(6)}_{1/2}(1,1,0)  
+3\,\mathcal{F}^{(6)}_{1/3}(2,1,0)
- 12 \mathcal{F}^{(6)}_{1/3}(1,1,0)-72 \right) \,,
\end{split}
\ee 
for $N=6$.  The various integrals can then be computed using the results of the previous section.
As an example, from the previous decompositions and from Eq. \eqref{finalintegral00} we get the following expressions for the ${\rm E}_8$ gauge threshold of the $\mathbb{Z}_2$ and $\mathbb{Z}_3$ orbifolds
\be
\begin{split}
\Delta_{{\rm E}_8}^{(2)}= \sum\limits_{p_{\rm L}^2-p_{\rm R}^2=4}  e^{2\pi \I \vec\lambda\cdot \vec p}&\left[1+\frac{p_{\rm R}^2}{4}\log\left(\frac{p_{\rm R}^2}{p_{\rm L}^2}\right)\right]
\\
-8\sum\limits_{p_{\rm L}^2-p_{\rm R}^2=2}&\left[1+\frac{p_{\rm R}^2}{2} \log\left(\frac{p_{\rm R}^2}{p_{\rm L}^2}\right)\right]-144\,\, {\rm Res}\, \mathcal{E}^{(2)\star}_2 \ar{0}{1} (U,T; s) \Big|_{s=1} +\textrm{const}\,.
\end{split}
\ee
\be
\begin{split}
\Delta_{{\rm E}_8}^{(3)}= \sum\limits_{p_{\rm L}^2-p_{\rm R}^2=4}&  e^{2\pi \I \vec\lambda\cdot \vec p}\left[1+\frac{p_{\rm R}^2}{4}\log\left(\frac{p_{\rm R}^2}{p_{\rm L}^2}\right)\right] -9\sum\limits_{p_{\rm L}^2-p_{\rm R}^2=4/3}\left[1+\frac{3 p_{\rm R}^2}{4} \log\left(\frac{p_{\rm R}^2}{p_{\rm L}^2}\right)\right]
\\
&-6\sum\limits_{p_{\rm L}^2-p_{\rm R}^2=8/3}\left[1+\frac{3 p_{\rm R}^2}{8} \log\left(\frac{p_{\rm R}^2}{p_{\rm L}^2}\right)\right]-144\,\, {\rm Res}\, \mathcal{E}^{(3)\star}_2 \ar{0}{1} (U,T; s) \Big|_{s=1}+\textrm{const}\,.
\end{split}
\ee
As stressed in the introduction, by unfolding $\cF_N$ against the Niebur-Poincar\'e series, the singularity structure of the amplitudes becomes crystal-clear and one may, for example, prove in a chamber-independent fashion that the above gauge thresholds are regular at any point of the Narain moduli space, as expected. The evaluation of $\Gamma_0(N)$-invariant modular integrals in the presence of Wilson lines and lattice momentum insertions proceeds in a similar fashion, see \cite{Angelantonj:2012gw}.


\subsection{Thermal type II: a very special example}

We shall conclude this section by applying our method to the evaluation of the free energy of  special type II $(4,0)$ theories in two dimensions. We shall focus our attention on the `Hybrid' thermal vacua constructed in \cite{Florakis:2010ty,Florakis:2010is}. These vacua are free of Hagedorn divergences and their right-moving supersymmetries are broken spontaneously at the string level and replaced by a \emph{Massive Spectral boson-fermion Degeneracy Symmetry} (MSDS) structure, studied in \cite{Kounnas:2008ft,Florakis:2009sm}. This degeneracy symmetry of the spectrum, which manifests itself only at special factorised points of the moduli space of string theories compactified to two dimensions, arises due to the presence of a chiral spectral-flow operator which is responsible for mapping the bosonic tower of states into the fermionic one, with the exception of the massless level. The resulting chiral spectrum, encoded in the standard vectorial and spinorial ${\rm SO}(24)$ characters, is then Bose-Fermi degenerate at all massive levels.

 The free energy density of these theories is given by the modular integral \cite{Florakis:2010ty}
\begin{equation}
F = \frac{R}{8\pi}\int_{\cF}\de\mu\, \frac{E_4\, (\bar{V}_{24}-\bar{S}_{24})}{\eta^{12}}\,\frac{1}{2}\sum\limits_{a,b=0}^{1}(-)^{a+b}\, \theta^4 \ar{a/2}{b/2} \,
\sum_{m,n\in\mathbb{Z}}e^{-\frac{\pi R^2}{\tau_2}|m+\tau n|^2}(-)^{ma+nb+mn}
\,,
\end{equation}
where the  temperature is identified with the inverse radius $R$ of the thermal time cycle and $V_{24}, S_{24}$ are the standard ${\rm SO} (24)$ characters associated to the vectorial and spinorial representations. As shown in \cite{Florakis:2010ty} the integrand can be cast in a $\mathbb{Z}_2$ orbifold description where the 
generator involves a momentum shift along the thermal radius accompanied by $(-1)^{F_{\rm L}}$. As a result, one obtains   
\begin{equation}
F=\frac{1}{8\pi}\int_{\cF_2}\de\mu\, \frac{\theta_2^4 \,E_4\, (\bar{V}_{24}-\bar{S}_{24})}{\eta^{12}}\, \varGamma_{(1,1)}[^0_1](R)\, .
\end{equation} 
This integral can now be evaluated using Eq. \eqref{finalintegral00}, in view of the remarkable identity $\bar{V}_{24}-\bar{S}_{24}=24$, which is a direct consequence of the MSDS structure, and of the identification
\begin{equation}
\frac{\theta_2^4\,E_4}{\eta^{12}} 
= \cF^{(2)}_{0}(1,1,0)\,.
\end{equation}
Therefore, one finds
\be
\begin{split}
F &= \frac{6}{\pi} \, {\mathcal I}_{1,1} (1,1;0)
\\
&= 
24\, \left(\,R+\frac{1}{2R} - \left|R-\frac{1}{2R}\right|\, \right)\,.
\end{split}
\label{MSDSconic}
\ee
This expression reproduces the free energy of \cite{Florakis:2010ty,Florakis:2010is} in a straightforward and  chamber-independent fashion. In particular, 
the T-duality symmetry $R\rightarrow (2R)^{-1}$ is manifest throughout, and Eq. \eqref{MSDSconic} clearly displays the conical singularity at $R=1/\sqrt{2}$, that 
signals the onset of a stringy phase transition in the `Hybrid' non-singular toy-model universe \cite{Florakis:2010is}.

\subsection*{Acknowledgements}

C.A. would like to thank the TH Unit at CERN, the Arnold Sommerfeld Centre at the Ludwig-Maximilians-Universit\"at M\"unchen and the Max-Planck-Institut f\"ur Physik in M\"unchen for hospitality during different stages of this project. I.F. would like to thank the TH Unit at CERN and the Physics Department at the University of Torino for hospitality during different stages of this project. B.P. is grateful to K. Bringmann for useful correspondence. This work was partially supported by the European ERC Advanced Grant no. 226455 ``Supersymmetry, Quantum Gravity and Gauge Fields'' (SUPERFIELDS), the ERC Advanced Grant 226371 ``MassTeV'', by the CNRS PICS Nos. 3747 and 4172, and by the Italian MIUR-PRIN contract 2009KHZKRX-007 ``Symmetries of the Universe and of the Fundamental Interactions''.


\appendix

\section{Kloosterman-Selberg zeta function for $\Gamma_0(N)$}\label{appendixKloosterman}

The Kloosterman-Selberg zeta function associated to a pair of cusps $\mathfrak{ab}$ of the
Hecke congruence subgroup $\Gamma_0(N)$ is defined for ${\rm Re}(s)>1$ by the absolutely convergent sum  
\be
\label{Zivgen}
\cZ_{\mathfrak{ab}} (m,n;s) = 
 \sum_{{\tiny\left( \begin{array}{cc} a & * \\ c & d \end{array}\right)} \in 
 \Gamma_\infty \backslash 
 \sigma_\mathfrak{a}^{-1}\, \Gamma_0(N) \, \sigma_\mathfrak{b} / \Gamma_\infty} 
 \!\!\!\!\!\!\!\! \!\!\!\!\!\!\!\!
 e^{2\I\pi ( m \frac{d}{c}+ n \frac{a}{c} )} \times
\begin{cases}
\frac{1}{2c\sqrt{|mn|}}
\, J_{2s-1} \left( \frac{4\pi}{c} 
\sqrt{mn} \right)  & \mbox{if}\quad mn>0\,,
\\
\frac{1}{2c\sqrt{|mn|}}
\, I_{2s-1} \left( \frac{4\pi}{c} 
\sqrt{-mn} \right) & \mbox{if}\quad mn<0 \,,
\\
\frac{1}{c^{2s}} & \mbox{if}\quad mn=0\ .
\end{cases}
\ee
The sum runs over $2\times 2$ real matrices $ {\tiny\left( \begin{array}{cc} a & * \\ c & d \end{array}\right)}$ in the double cosets $ \Gamma_\infty \backslash 
 \sigma_\mathfrak{a}^{-1}\, \Gamma_0(N) \, \sigma_\mathfrak{b} / \Gamma_\infty$, where
$\sigma_\mathfrak{a}$ is the {\em scaling matrix} associated to the cusp $\mathfrak{a}$, and is defined in  \eqref{defscalmat}. In this
Appendix, we provide a more explicit expression for the general Kloosterman-Selberg zeta function
$\cZ_{\mathfrak{ab}} (m,n;s)$, and evaluate it in terms of the Riemann zeta function
in the special case where $mn=0$.

\subsection{$\Gamma_0(N)$ with $N$ prime}

For $N$ prime, using
\eqref{sigmaNprime}, $\sigma_\mathfrak{a}^{-1}\, \Gamma_0(N) \, \sigma_\mathfrak{b}$ can be parameterised by
\begin{equation}
\label{doublecosetN}
\begin{split}
\sigma_\infty^{-1} \,\lambda \, \sigma_\infty &=
\left(\begin{array}{cc} a & b \\ c & d \end{array}\right) = 
 \left(\begin{array}{cc} \alpha & \beta \\ N\gamma & \delta \end{array}\right) \,,
\\
\sigma_0^{-1} \,\lambda \, \sigma_0 &=
\left(\begin{array}{cc} a & b \\ c & d \end{array}\right) = 
 \left(\begin{array}{cc} \alpha & \beta \\ N\gamma & \delta \end{array}\right) \,, 
\\
\sigma_\infty^{-1} \,\lambda \, \sigma_0 &=
\left(\begin{array}{cc} a & b \\ c & d \end{array}\right) = 
\left(\begin{array}{cc} \sqrt{N} \alpha & \beta/\sqrt{N} \\ \sqrt{N}\gamma & \sqrt{N} \delta \end{array}\right) \,,
\end{split}
\end{equation}
where $\lambda={\tiny \begin{pmatrix} \alpha & \beta \\ N \gamma & \delta \end{pmatrix} }\in \Gamma_0 (N)$.
Using this parameterisation, one may rewrite for instance Eq. \eqref{Zivgen} as
\be
\label{Zivgenexp}
\cZ_{\infty\infty} (m,n;s) = 
\sum_{\substack{c>0\\c=0\,\mod N}} \sum_{d\in (\mathbb{Z}/c\mathbb{Z})^*} \exp\left[ \tfrac{2\pi \I}{c} (m \, d+ n\, d^{-1})\right]\ \, \times
\begin{cases}
\frac{1}{2c}\sqrt{\frac{1}{|mn|}}
\, J_{2s-1} \left( \frac{4\pi}{c} 
\sqrt{mn}\right)  \\
\frac{1}{2c}\sqrt{\frac{1}{|mn|}}
\, I_{2s-1} \left( \frac{4\pi}{c} 
\sqrt{-mn}\right) 
\\
\frac{1}{c^{2s}} 
\end{cases}\,,
\ee
and
\be
\label{Zivgenexp2}
\cZ_{0\infty}(m,n;s) = \sum_{c>0}
\sum_{\substack{0\leq d<c\\ (c,Nd)=1}}
 \exp\left[ \tfrac{2\pi \I}{c} ( m d + n (N d)^{-1}) \right]
 \times
\begin{cases}
\frac{1}{2c\sqrt{N|mn|}}\, J_{2s-1} \left( \frac{4\pi}{c} \sqrt{\frac{mn}{N}}\right) 
\\
\frac{1}{2c\sqrt{N|mn|}}\, I_{2s-1} \left( \frac{4\pi}{c} \sqrt{\frac{-mn}{N}}\right)  \,.
\\
\frac{1}{N^s\, c^{2s}} 
\end{cases}
\ee

For $mn=0$, the Kloosterman-Selberg zeta function $\cZ_{\mathfrak{ab}} (m,n;s)$ can be evaluated
in terms of the Riemann zeta function. For $N$ prime one finds
\begin{equation}
\label{ZNS0}
\begin{split}
\cZ_{\infty\infty}(0,0;s) &=\frac{N-1}{N^{2s}-1} \frac{\zeta (2s-1)}{\zeta (2s)}\,,
\\
 \cZ_{\infty 0}(0,0;s) &= \frac{N^{2s-1}-1}{N^{s-1}(N^{2s}-1)}\frac{\zeta(2s-1)}{\zeta(2s)} \,,
 \end{split}
 \end{equation}
 and, for $m\not=0$
 \begin{equation}
 \label{ZNS0bis}
\begin{split} 
\cZ_{\infty\infty}(0,\pm m;s) &= \frac{N \sigma_{1-2s} (m/N )-\sigma_{1-2s} (m )}{(N^{2s}-1) \zeta (2s)}\,,
\\
\cZ_{\infty 0}(0,\pm m;s) &= \frac{N^{2s-1} \sigma_{1-2s} (m)-\sigma_{1-2s} (m/N)}{N^{s-1}(N^{2s}-1)\zeta (2s)}\,,
\end{split}
\ee
where $\sigma_{t} (n)$ is the divisor function, and it is understood that $\sigma_{1-2s} (m/N)$ vanishes unless $N$ divides $m$. These expressions can be derived either by direct evaluation of Eqs. \eqref{Zivgenexp} and \eqref{Zivgenexp2}, or by using the relation \eqref{defvarphi} between  non-holomorphic Eisenstein series of $\Gamma_0 (N)$ and non-holomorphic Eisenstein series
 of  the full modular group $\Gamma$, since their Fourier coefficients \eqref{fourier} and \eqref{zeromode} are related to the Kloosterman-Selberg zeta functions with $mn=0$.

\subsection{$\Gamma_0(4)$}

For $N=4$, aside from the cusps at $0$ and $\infty$ with scaling matrices \eqref{sigmaNprime}, there is an additional cusp at the rational point $\frac{1}{2}$ with width $m_{1/2}=1$, and scaling matrix
\begin{equation}
\sigma_{1/2} = \tau_{1/2} =  \left( \begin{array}{cc} 1 & 0 \\ 2 & 1 \end{array}\right)\,.
\end{equation}
The double-cosets entering the definition of the Kloosterman-Selberg zeta function can be obtained by conjugating a generic element of $\Gamma_0 (4)$ by these matrices and those in Eq. \eqref{sigmaNprime}. For instance, if $\lambda ={\tiny \begin{pmatrix} \alpha & \beta \\ 4 \gamma & \delta \end{pmatrix}} \in \Gamma_0 (4)$ then
\begin{equation}
\sigma_{\infty}^{-1}\, \lambda\, \sigma_{1/2} 
=
\left(\begin{array}{cc} a & b \\ c & d \end{array}\right)  = \begin{pmatrix} \alpha +2\beta & \beta
\\
4\gamma +2\delta  & \delta\end{pmatrix} \,,
\end{equation}
and similarly for the other combinations of cusps. 

In order to compute the Kloosterman-Selberg zeta functions for $mn=0$, we use the fact that 
the non-holomorphic Eisenstein series \eqref{eisenstein} can be expressed
as linear combinations of Eisenstein series under the full modular group
\be
 \begin{split}
\label{relation4}
E^{(4)\star}_\infty(\tau ,s) =&  \frac{E^\star(4\tau,s) - 2^{-s} E^\star(2\tau,s)}{4^{s}-1} \,,
\\
E^{(4)\star}_0(\tau,s) =&  \frac{E^\star(\tau,s) - 2^{-s} E^\star(2\tau,s)}{4^{s}-1} \,,
\\
E^{(4)\star}_{1/2}(\tau,s) = & \frac{(2^s+2^{-s})E^\star(2\tau,s) - E^\star(\tau,s)-E^\star(4\tau,s)}{4^{s}-1} \,.
\end{split}
\ee
Using the 
functional equation for the Eisenstein series for the full modular group, we find, in the same basis,
\be
\varPhi(s)=\frac{1}{2^{2s}-1}
\begin{pmatrix}
2^{1-2s} & 1-2^{1-2s} &  1-2^{1-2s}
\\
1-2^{1-2s} & 2^{1-2s} &  1-2^{1-2s}
 \\
1-2^{1-2s} &  1-2^{1-2s} & 2^{1-2s}
\end{pmatrix}\, .
\ee
It follows that the Kloosterman-Selberg zeta function for $m=n=0$ is given by 
\begin{equation}
\begin{split}
\cZ^{(4)}_{\infty\infty}(0,0;s)= & \frac{2^{1-2s}}{ 2^{2s}-1} \, \frac{\zeta (2s-1)}{\zeta (2s)}\,,
\\
\cZ^{(4)}_{\infty 0}(0,0;s)= & \frac{1-2^{1-2s}}{2^{2s}-1}\, \frac{\zeta (2s-1)}{\zeta (2s)}\,,
\\
\cZ^{(4)}_{\infty \frac{1}{2}} (0,0;s) =& \frac{1-2^{1-2s}}{2^{2s}-1} \, \frac{\zeta (2s-1)}{\zeta (2s)}
\,.
\end{split}
\ee
For $n=0$ and $m\neq 0$, we have instead
\begin{equation}
\begin{split}
\cZ^{(4)}_{\infty\infty}(0,\pm m;s) = &\frac{4\sigma_{1-2s} (m/4 )- 2 \sigma_{1-2s} (m/2 )}{2^{2s}(2^{2s}-1)\zeta (2s)}\,,
\\
\cZ^{(4)}_{\infty 0}(0,\pm m ;s) = &\frac{\sigma_{1-2s} (m )- 2^{1-2s} \sigma_{1-2s} (m/2)} {(2^{2s}-1)\zeta (2s)} \,,
\\
\cZ^{(4)}_{\infty \frac{1}{2}}(0,\pm m;s) 
= &  \frac{ 2(1+2^{-2s}) \sigma_{1-2s}(n/2) - \sigma_{1-2s}(n) - 2^{2-2s} \sigma_{1-2s} (n/4) } { (2^{2s}-1) \zeta(2s) }\,.
\end{split}
\ee
As before, it is understood that $\sigma_{1-2s} (m/M)$ vanishes unless $M$ divides $m$. The remaining Kloosterman-Selberg zeta functions (with $mn=0$) can be obtained by symmetry and/or by use of the relations \eqref{conjugaterelation}.

\subsection{$\Gamma_0(6)$}

For $N=6$, aside from the cusps at $0$ and $\infty$ with scaling matrices \eqref{sigmaNprime}, there are two additional (inequivalent) cusps at the rational points $\frac{1}{2}$ and $\frac{1}{3}$, of width 3 and 2, respectively. The associated $\tau$ and $\sigma$ matrices can be chosen to be
\begin{equation}
\begin{split}
\tau_{1/2} =& \left( \begin{array}{cc} 1 & -2 \\ 2 & -3 \end{array} \right)\,, \quad m_{1/2}=3\,,\quad
\sigma_{1/2} = \left( \begin{array}{cc} \sqrt{3} & -2/\sqrt{3} \\ 2\sqrt{3} & -\sqrt{3} \end{array} \right)\,,
\\
\qquad
\tau_{1/3} =& \left( \begin{array}{cc} 1 & -1 \\ 3 & -2 \end{array} \right)\,, \quad m_{1/3} =2\,,\quad
\sigma_{1/3} = \left( \begin{array}{cc} \sqrt{2} & -1/\sqrt{2} \\ 3\sqrt{2} & -\sqrt{2} \end{array} \right)\,.
\end{split}
\end{equation}
The double-cosets entering the definition of the Kloosterman-Selberg zeta function can be obtained by conjugating a generic element of $\Gamma_0 (6)$ by these matrices and those in Eq. \eqref{sigmaNprime}. For instance, if $\lambda ={\tiny \begin{pmatrix} \alpha & \beta \\ 6 \gamma & \delta \end{pmatrix}} \in \Gamma_0 (6)$ then
\begin{equation}
\begin{split}
\sigma_{\infty}^{-1}\, \lambda\, \sigma_{1/2} 
&
=
\left(\begin{array}{cc} a & b \\ c & d \end{array}\right) = \begin{pmatrix} \sqrt{3} (\alpha +2\beta) & -(2\alpha +3\beta )/\sqrt{3}\\
2\sqrt{3} (3\gamma +\delta ) & -\sqrt{3} (4\gamma +\delta )\end{pmatrix} \,,
\\
\sigma_{\infty}^{-1}\, \lambda\, \sigma_{1/3} 
&
=
\left(\begin{array}{cc} a & b \\ c & d \end{array}\right) = \begin{pmatrix} \sqrt{2} (\alpha +3\beta) & -(\alpha +2\beta )/\sqrt{2}\\
3\sqrt{2} (2\gamma +\delta ) & -\sqrt{2} (3\gamma +\delta )\end{pmatrix} \,,
\end{split}
\end{equation}
and similarly for the other combinations of cusps.

The non-holomorphic Eisenstein series of $\Gamma_0 (6)$ have the following decomposition in terms of
linear combinations of Eisenstein series of the full modular group
\be
\begin{split}
E^{(6)\star}_\infty(s,\tau) = & \frac{6^s E^\star(s,6\tau)- 3^s E^\star(s,3\tau)-2^s E^\star(s,2\tau)+E^\star(s,\tau)}{(1-4^s)(1-9^s)}\,,
\\
E^{(6)\star}_0(\tau,s) =&  \frac{6^s E^\star(\tau,s)- 3^s E^\star(2\tau,s)-2^s E^\star(3\tau,s)+E^\star(6\tau,s)}{(1-4^s)(1-9^s)}\,, 
\\
E^{(6)\star}_{1/3}(\tau,s) =&  \frac{6^s E^\star(3\tau,s)- 3^s E^\star(6\tau,s)-2^s E^\star(\tau,s)+E^\star(2\tau,s)}{(1-4^s)(1-9^s)} \,,
\\
E^{(6)\star}_{1/2}(\tau,s) =&  \frac{6^s E^\star(2\tau,s)- 3^s E^\star(\tau,s)-2^s E^\star(6\tau,s)+E^\star(3\tau,s)}{(1-4^s)(1-9^s)} \,.
\end{split}
\label{relation6}
\ee
Using the functional equation for the Eisenstein series for the full modular group and the decomposition \eqref{relation6}, or explicitly Eqs. \eqref{scattering} and \eqref{scattering2}, we find, in the same basis,
\be
\varPhi(s)=
 \frac{1}{(4^s-1)(9^s-1)}
\begin{pmatrix}
2 & \alpha_2 \, \alpha_3 & 2\, \alpha_2 & \alpha_3  
\\
 \alpha_2 \, \alpha_3 & 2 &  \alpha_3 & 2\, \alpha_2  
 \\
 2\, \alpha_2& \alpha_3 & 2 &   \alpha_2\, \alpha_3 
 \\
 \alpha_3 & 2\, \alpha_2  &  \alpha_2 \, \alpha_3  & 2
\end{pmatrix} \,,
\ee
where $\alpha_p = p^s - p^{1-s}$.
The Kloosterman-Selberg zeta functions for $m=n=0$ are easily read off from $\varPhi(s)$,
\be
\begin{split}
\cZ^{(6)}_{\infty\infty}(0,0;s)=&\frac{2}{(2^{2s}-1)(3^{2s}-1)}\,
 \frac{\zeta (2s-1)}{\zeta (2s)}\,,
\\
\cZ^{(6)}_{\infty0}(0,0;s)=&\frac{6^{1-s}(2^{2s-1}-1)(3^{2s-1}-1)}
{(2^{2s}-1)(3^{2s}-1)}\,
 \frac{\zeta (2s-1)}{\zeta (2s)}\,,
\\
\cZ^{(6)}_{\infty\frac{1}{3}}(0,0;s)=&\frac{2^{1+s}-2^{2-s}}{(2^{2s}-1)(3^{2s}-1)}\,
 \frac{\zeta (2s-1)}{\zeta (2s)}\,,
\\
\cZ^{(6)}_{\infty\frac{1}{2}}(0,0;s)=&\frac{3^s-3^{1-s}}
{(2^{2s}-1)(3^{2s}-1)}\,\frac{\zeta (2s-1)}{\zeta (2s)}\,.
\end{split}
\ee
For $m\not=0$, we have instead
\be
\begin{split}
\cZ^{(6)}_{\infty\infty}(0,\pm m;s) 
=& \frac{6\, \sigma_{1-2s} (m/6 )- 2\, \sigma_{1-2s} (m/2 )
- 3 \,\sigma_{1-2s} (m/3 )+ \sigma_{1-2s} (m)}{(2^{2s}-1)(3^{2s}-1)\,\zeta (2s)}\,,
\\
\cZ^{(6)}_{\infty 0}(0,\pm m;s) =& \frac{6^{s}\, \sigma_{1-2s} (m)
- 2^{s} 3^{1-s}\, \sigma_{1-2s} (m/3 )
- 3^{s} 2^{1-s}\, \sigma_{1-2s} (m/2)+
6^{1-s} \, \sigma_{1-2s} (m/6)}{(2^{2s}-1)(3^{2s}-1)\,\zeta (2s)}\,,
\\
\cZ^{(6)}_{\infty\frac{1}{3}}(0,\pm m;s) =& \frac{6^s 3^{1-s} \, \sigma_{1-2s} (m /3)
- 3^s 6^{1-s} \, \sigma_{1-2s} (m/6)
- 2^s \, \sigma_{1-2s} (m )
+ 2^{1-s}\, \sigma_{1-2s} (m/2)}
{(2^{2s}-1)(3^{2s}-1)\, \zeta (2s)}\,,
\\
\cZ^{(6)}_{\infty\frac{1}{2}}(0,\pm m;s) =& \frac{6^{s}2^{1-s}\sigma_{1-2s} (m/2)
- 3^{s} \sigma_{1-2s} (m )
- 2^{s} 6^{1-s} \sigma_{1-2s} (m/6)+
3^{1-s} \sigma_{1-2s} (m/3)}{(2^{2s}-1)(3^{2s}-1)\zeta (2s)}\,.
\end{split}
\end{equation}
As before,  it is understood that $\sigma_{1-2s} (m/M)$ vanishes unless $M$ divides $m$. The remaining Kloosterman-Selberg zeta functions (with $mn=0$) can be obtained by symmetry and/or by use of the relations \eqref{conjugaterelation}.

\section{A compendium on modular forms for $\Gamma_0(N)$ 
\label{sec_comp}}

In this Appendix we collect some standard facts about holomorphic modular forms under the congruence
subgroups $\Gamma_0(N)$, with special focus on the values $N=2,3,4,6$ relevant for orbifold
string compactifications. Most of these facts can be found in \cite{shimura,iwaniec1,iwaniec2,0791.11022,0804.11039,0697.10023,Zagier123,1188.11074,webrefs}.

\subsection{Generalities}\label{B1generalities}

For any integer $N$ and even\footnote{We restrict attention to the case of even weight in order to simplify the presentation. More general cases require the introduction of suitable multiplier systems \cite{shimura,iwaniec1,iwaniec2}.} integer $w$, we denote by  $\cM_{w} (N)$  the space of holomorphic modular forms of weight $w$ under the Hecke congruence subgroup 
$\Gamma_0(N)$. Those are defined by the condition of covariance under $\Gamma_0 (N)$,
\begin{equation}
f (\tau) \, \big\vert_{w} \, \gamma = f(\tau) \,,\qquad   {\rm for\ any}\quad \gamma = \begin{pmatrix} a & b \\ c & d\end{pmatrix}\in\Gamma_0 (N)\,,
\label{ABdefmodular}
\end{equation}
together with the condition of holomorphy in $\mathbb{H}$, in particular at each of the cusps
of  $\Gamma_0 (N)$. In \eqref{ABdefmodular} the Petersson slash operator is defined by:
\begin{equation}
	f (\tau) \, \big\vert_{w} \, \gamma \equiv  j_\gamma^{-w} (\tau ) \, f (\gamma \tau) \,,\qquad j_\gamma (\tau ) = c\, \tau + d \,.
\end{equation}
Since each cusp $\mathfrak{a}$ can be mapped to $\infty$
by a scaling matrix $\sigma_\mathfrak{a}$, holomorphy is tantamount to requiring that the Fourier expansion 
of $f$ at each cusp $\mathfrak{a}$ takes the form
\begin{equation}
\label{FourierHol}
f (\tau) \big|_w \sigma_\mathfrak{a} = \sum_{n=0}^\infty \tilde f_\mathfrak{a} (n)\, q^n\ ,\qquad 
q\equiv e^{2\I\pi \tau}\ ,
\end{equation}
so that $f (\tau) \big| \sigma_\mathfrak{a}$ is finite at $q=0$. The space  of holomorphic cusp forms  $\cS_{w} (N) \subset \cM_{w} (N)$ is defined by the stronger condition $\tilde f _\mathfrak{a} (0) = 0$ at all cusps. The space of weak holomorphic modular forms $\cM_{w}^{!} (N)$ is defined by 
the weaker condition of meromorphy in $\mathbb{H}$, with poles only at the cusps. In pratice,
it means that $f(\tau)$ has a Fourier expansion at the cusp $\mathfrak{a}$ of the form
\begin{equation}
\label{FourierweakHol}
f (\tau) \big|_w \sigma_\mathfrak{a} = \sum_{n=-\kappa_a}^\infty \tilde f_\mathfrak{a} (n)\, q^n\ ,\end{equation}
for some positive integer $\kappa_a$. 

The direct sum $\cM_\star (N)=\bigoplus_{w\geq 0} \cM_w (N)$ forms a graded algebra under the usual product, and so does $\cS_\star(N)$, while $\cM^{!}_\star(N)$ is a graded field. Since modular forms for the full modular group are also modular under $\Gamma_0(N)$, $\cM_\star (N)$ is a module over $\cM_\star (1)$. The latter is well-known to be a polynomial ring generated by the Eisenstein series $E_4$ and $E_6$, where 
\be
\label{defeis}
E_{w} (\tau ) = 1+ \frac{2}{\zeta (1-w)} \sum_{n=1}^\infty \sigma_{w-1} (n)\, q^n\, .
\ee
In particular, $\varDelta = \frac{1}{1728}\,(E_4^3 - E_6^2)$ is the lowest weight cusp form for the
full modular group.
For values of $N$ such that the modular curve $\Gamma_0(N)\backslash (\mathbb{H} \cup \mathbb{Q} \cup \{\infty\})$ has genus zero (which is the case for all $N\leq 10$, including the cases $N=1,2,3,4,6$ of interest  in this work), $\cM_\star (N)$ is a free module over $\cM_\star (1)$ generated by $\nu_N$ generators, where $\nu_N$ is the index of $\Gamma_0(N)$ inside ${\rm SL}(2;\IZ)$ \cite{0791.11022}. Under the same condition, $\cM^{!}_0(N)$ is isomorphic to the field of rational functions in one variable $J_N$, defined uniquely up to M\"obius transformations. The unique choice such that $J_N(\tau)=1/q+\cO(q)$ at the cusp at infinity is known as the Hauptmodul. Modular curves of genus 0 are famously related to  {\em monstrous moonshine}, with the Hauptmodul being interpreted as 
a McKay-Thomson series associated to a certain conjugacy class of the Monster group \cite{conway}.

An important class of modular forms for $\Gamma_0(N)$ arises from modular forms $f_{d}$ of 
$\Gamma_0(d)$ for any divisor $d$ of $N$, via $f_N(\tau)=f_d(N\tau/d)$. In particular, if $f(\tau)$ is a modular form for the full modular group, $f(N\tau)$ is a modular form for $\Gamma_0(N)$. In particular, even though $E_2(\tau)$ defined by \eqref{defeis} is {\it not} a modular form 
under ${\rm SL}(2;\IZ)$,
\begin{equation}
\label{defXN}
X_N (\tau ) \equiv  E_2 - N \, E_2 (N \tau ) = \hat E_2 - N \, \hat E_2 (N \tau )
\end{equation}
is a holomorphic modular form of $\Gamma_0(N)$ of weight 2. Indeed, its covariance under $\Gamma_0(N)$  follows from the covariance of $\hat E_2\equiv E_2-\frac{3}{\pi\tau_2}$ under 
${\rm SL}(2;\IZ)$. In \eqref{defXN} and elsewhere, an Eisenstein series $E_w$ (or any other modular form) without explicit argument denotes $E_w\equiv E_w(\tau)$. 

Just as for the full modular group, we define the space of weakly almost holomorphic modular forms $\cM^\times_w (N)$ to be the 
weight $w$ subspace of the algebra of polynomials in $\hat E_2$ with coefficients in $\cM^{!}_\star(N)$. This algebra admits an action of
the modular derivative operator $D$ defined in \eqref{modularderiv}, which maps  $\cM^\times_{w} (N)$ to $\cM^\times_{w+2}(N)$. Its action on 
the (almost) holomorphic Eisenstein series is given by 
\be
D\, \hat E_2 = \tfrac{1}{6} (E_4 - \hat E_2^2 )\,, \qquad
D\, E_{4}  =\tfrac{2}{3} (E_6 - \hat E_2 \, E_4 )\,, \qquad D \, E_6 = E_4^2 - \hat E_2 \, E_6\,,
\ee
from which it easily follows that
\be
D\, X_N 
=\tfrac{1}{6} (E_4 - N^2 \, E_4(N\tau) + X_N^2) - \frac13 \hat E_2\, X_N\ .
\ee

Having recalled these general facts, we now discuss the cases $N=2,3,4,6$ relevant for this work in some more detail.

\subsection{$\Gamma_0(2)$  \label{appg02}}

The congruence subgroup $\Gamma_0(2)$ has index 3 in ${\rm SL}(2;\IZ)$ and 2 cusps at $\infty$ and $0$. Its fundamental domain can be chosen as $\cF_2 = \{1,S,ST\}\cF$. The dimensions $d_w^{(2)}$ of $\cM_w(2)$ are given by the generating function
\be
\label{dimG2}
\sum_{k\ge 0} d_{2k}^{ (2)}\, x^{2k} 
= \frac{1+x^2+x^4}{(1-x^4)(1-x^6)} = 1+x^2+2x^4+2 x^6+3 x^8 + 3 x^{10}+\dots
\ee
The ring of holomorphic modular forms is generated by the two elements $X_2(\tau)$ and $E_4(2\tau)$. In accordance with \eqref{dimG2}, 
any element of $\cM_\star(2)$ can be decomposed uniquely as 
\be
\phi = A + B \, X_2 + C \, X_2^2
\ee 
where $A,B,C$ are modular forms of ${\rm SL}(2;\IZ)$.  For instance,
\be
E_4(2\tau)=\tfrac14 (5\, X_2^2-E_4)\,,\quad 
X_2^3 = \tfrac{1}{4} \left( 3 \, X_2\, E_4 - E_6 \right) \,.
\ee
The first cusp form occurs at weight 8 and is given by
\be
\begin{split}
\varDelta_8^{(2)}&= [\eta(\tau)\eta(2\tau)]^8 
=  -\tfrac{1}{144}\, X_2^4 + \tfrac{5}{576}\, X_2^2 \, E_4 - \tfrac{1}{576}\, E_4^2 
\\
&=  q - 8 \, q^2+12\, q^3+64\, q^4-210\, q^5+ \cO(q^6)\,.
\end{split}
\ee
To derive the behaviour near the cusp 0, it suffices to determine the action of the scaling matrix $\sigma_0$ on the generators:
\begin{equation}
X_2  \big|\sigma_0 =- X_2 \,,\quad
E_4 \big|\sigma_0 = 4 \, E_4 (2\tau ) \,, \quad
\hat E_2 \big| \sigma_0 = \hat E_2 -X_2\,.
\end{equation}
In particular, the cusp form $\varDelta_8^{(2)}$ is even under the action of $\sigma_0$.

The Hauptmodul for $\Gamma_0(2)$ is the McKay-Thompson series associated to the conjugacy class $2B$ of the Monster,
\be
J_2 = \frac{\varDelta(\tau)}{\varDelta(2\tau)}+24 = \frac{48\, X_2^2}{X_2^2 - E_4 (2\tau )}-40 = 
q^{-1} + 276 \, q-2048\, q^2 + 11202\, q^3 + \cO (q^4)\,.
\ee
It has a simple pole at the cusp $\infty$ and is regular at the other cusp 0,
\be
J_2 \big|\,\sigma_0 =  \frac{4096}{J_2-24}+24= 24 + 4096 \, q + 98304 \, q^2 + 1228800\,  q^3 +\cO(q^4)
\,.
\ee
The Hauptmodul of ${\rm SL} (2;\IZ)$ can be expressed in terms of $J_2$ as
\be
J = J_2+\frac{196608}{J_2-24}+\frac{16777216}{(J_2-24)^2}\,.
\ee
Modular derivatives act as 
\be
\begin{split}
D X_2 = & -\tfrac13 \, X_2 \, \hat E_2 + \tfrac13 \, X_2^2 -
 \tfrac43 \, E_4 (2\tau ) \,,
\\
D E_4 (2\tau ) =& - \tfrac{2}{3}\, \hat E_2 \, E_4 (2\tau ) -\tfrac{4}{3}\, X_2\, E_4 (2\tau) - X_2^3 
\,,
\\
D [1/\varDelta_8^{(2)}] = & ( 4 \, \hat E_2 - 2\,  X_2)/ (3 \, \varDelta_8^{(2)})\,.
\end{split}
\ee

\subsection{$\Gamma_0(3)$  \label{appg03}} 

The congruence subgroup $\Gamma_0(3)$ has index 4 in ${\rm SL}(2;\IZ)$ and 2 cusps at $\infty$ and $0$.  Its fundamental domain can be chosen as $\cF_3 = \{1,S,ST,ST^2\}\cF$. The dimensions $d^{(3)}_w$ of $\cM_w (3)$ are given by the generating function
\be
\sum_{k\ge 0}d_{2k}^{(3)}\, x^{2k} 
= \frac{1+x^2+x^4+x^6}{(1-x^4)(1-x^6)} = 1+x^2+2x^4+3 x^6+3 x^8 + 4 x^{10}+\dots  \label{dims03}
\ee
and the ring of holomorphic modular forms is generated by  $X_3(\tau)$, $E_4(3\tau)$ and by the unique cusp form of weight 6
\be
\begin{split}
\varDelta_6^{(3)} = & [\eta(\tau) \eta(3\tau)]^6 =  \tfrac{1}{384}\, X_3^3-\tfrac{7}{864}\, X_3 \, E_4+\tfrac{1}{216}\, E_6\\
 = & q - 6 \, q^2+9\, q^3+4\, q^4+6\, q^5-54 \, q^6+\cO(q^7)\,.
\end{split}
\ee 
In accordance with  \eqref{dims03}, any element of $\cM_\star(3)$ can be decomposed uniquely as 
\be
\phi = A + B \, X_3 + C\, X_3^2 + D\, X_3^3 \,,
\ee 
where $A,B,C,D$ are holomorphic modular forms of ${\rm SL}(2;\IZ)$, for instance
\be
\begin{split}
E_4(3\tau) &=-\tfrac19 \, E_4 + \tfrac{5}{18} \, X_3^2 \,, 
\\
E_6(3\tau)&=-\tfrac{1}{27} \, E_6 +\tfrac{7}{54}\, X_3 \, E_4 -\tfrac{35}{216}\, X_3^3\,,
\\
X_3^4 (\tau ) &= \tfrac{16}{27} \,  E_4^2 - \tfrac{64}{27}\, E_6 \, X_3 + \tfrac{8}{3} \, E_4\, X_3^2 \,.
\end{split}
\ee
To derive the behaviour near the cusp 0 it suffices to determine the action of the scaling matrix $\sigma_0$ on the generators:
\be
\begin{split}
X_3\big|\, \sigma_0 = - X_3 \,,\quad
E_4 (3 \tau ) \big|\,\sigma_0 =  \tfrac{5}{18}\, X_3^2 - E_4 (3\tau )
 \,, \quad
\hat E_2 \big| \, \sigma_0 = \hat E_2 - X_3\,.
\end{split}
\ee
In particular the cusp form $\varDelta_6 ^{(3)}$  is odd under $\sigma_0$.

 The Hauptmodul of $\Gamma_0(3)$ is the McKay-Thompson series associated to the conjugacy class $3B$ of the Monster
\be
\begin{split}
J_3=&\left(\frac{\eta(\tau)}{\eta(3\tau)}\right)^{12}+12 = \frac{X_3^3 - 36\, X_3 \, E_4 (3\tau ) - 960\, \varDelta_6^{(3)}}{64\, \varDelta_6^{(3)}}
\\
=& q^{-1}+54 \, q-76 \, q^2-243\, q^3+1188\, q^4 + \cO(q^5) \,.
\end{split}
\ee
It has a simple pole at the cusp $\infty$ and is regular at the other cusp, 
\be
 J_3 \big|\sigma_0= \frac{729}{J_3-12}+12
 = 12 + 729\, q + 8748\, q^2 + 65610\, q^3 +\cO(q^4)
\ee
The Hauptmodul of ${\rm SL} (2;\IZ)$ can be expressed in terms of $J_3$ as
\be
\begin{split}
J = & J_3+\frac{196830}{J_3-12}+\frac{19131876}{(J_3-12)^2}+\frac{387420489}{(J_3-12)^3}\,.
\end{split}
\ee
Modular derivatives act as 
\be
\begin{split}
D X_3 =& -\tfrac13 \, X_3 \, \hat E_2 -3 \, E_4 (3\tau ) - \tfrac{7}{12} \,X_3^2 \,,
\\
D E_4 (3\tau ) =&-\tfrac{2}{3}\, E_4 (3\tau )\, \hat E_2 -\tfrac{1}{2} X_3 \, E_4 (3\tau ) -\tfrac{131}{216} X_3^3 - 16\, \varDelta_6^{(3)}
\,,
\\
D [1/\varDelta_6^{(3)}] = & \frac{ \hat E_2 - \tfrac12 \, X_3}{ \varDelta_6^{(3)}}\,.
\end{split}
\ee

\subsection{$\Gamma_0(4)$ \label{appg04}}

The congruence subgroup $\Gamma_0(4)$ has index 6 in ${\rm SL} (2;\IZ)$ and 3 cusps at $\infty$, $0$ and $1/2$. It is 
isomorphic to the principal subgroup $\Gamma(2)$ under $\tau\to 2\tau$. Its fundamental domain can be chosen as $\cF_4 =\{ 1 ,S,ST,ST^2,ST^3,ST^2 S \}\, \cF$.
The dimensions $d^{(4)}_{w}$ of $\cM_w (4)$ are given by the generating function
\be
\sum_{k\ge 0} d_{2k}^{(4)}\, x^{2k} = \frac{1+2x^2+2x^4+x^6}{(1-x^4)(1-x^6)}=\frac{1}{(1-x^2)^2} = 1+2x^2+3x^4+4x^6+5x^8+\dots\,. \label{dims04}
\ee
The ring of holomorphic modular forms is generated by the weight 2 elements
\be
V_1 = X_2(2\tau)\,,\quad
V_2 = 
-\tfrac23\, \left[ X_4 -3 \, X_2 (2\tau ) \right]
\,.
\ee
In accordance with \eqref{dims04}, any element of $\cM_k(4)$ can be decomposed uniquely as 
\be
\phi = A + B_1 \, V_1 +B_2\, V_2 + C_1\, V_1^2 +C_2 \, V_2 ^2 + D\, V_1^3 \,,
\ee 
where $A,B_i ,C_i ,D$ are holomorphic modular forms of ${\rm SL}(2;\IZ)$.
We note the relations
\be
\begin{split}
\theta_2 ^4 (2\tau)  &= V_2\,, \qquad
\theta_3^4(2\tau) =\tfrac{1}{2}\, V_2 - V_1\ ,\qquad
\theta_4^4(2\tau) =-\tfrac12 \, V_2 - V_1\, ,
\\
X_2(\tau) &= V_1 - \tfrac32 \, V_2\,.
\end{split}
\ee
The first cusp form occurs at weight 6 and is given by
\be
\begin{split}
\varDelta_6^{(4)} =& \eta^{12} (2\tau) = \tfrac{1}{16}\, V_1^2\, V_2
 - \tfrac{1}{64}\, V_2^3 
\\
=&q-12\, q^3+54 \, q^5-88\, q^7 - 99\, q^9 + \cO(q^{10}) \,.
\end{split}
\ee
The scaling matrices $\sigma_0$ and $\sigma_{1/2}$ associated to the cusps $0$ and $1/2$ act on the generators as
\be
\begin{array}{lllll}
V_1 \big|\, \sigma_0 &=\tfrac{3}{4}\, V_2 -\tfrac{1}{2} \, V_1 \,,& \hspace*{1cm} &
V_1 \big|\, \sigma_{1/2} &= - \tfrac{3}{4} \, V_2 -\tfrac{1}{2}\, V_1 \,,
\\
V_2 \big|\, \sigma_0 &= \tfrac{1}{2}\, V_2 + V_1 \,, & &
V_2 \big|\, \sigma_{1/2} &= \tfrac{1}{2} \, V_2 - V_1 \,,
\\
\hat E_2 \big|\, \sigma_0 &= \hat E_2 + \tfrac{3}{2}\,  V_2 - 3 \, V_1 \,, & &
\hat E_2 \big|\, \sigma_{1/2} &= \hat E_2\,.
\end{array}
\ee
As a result, the weight-6 cusp form is odd under both $\sigma_0$ and $\sigma_{1/2}$.

The Hauptmodul for $\Gamma_0(4)$ is the McKay-Thompson series associated to the conjugacy class $4C$ of the Monster,
\be
\begin{split}
J_4  & 
= \left(\frac{\eta(\tau)}{\eta(4\tau)}\right)^8 + 8
= \frac{V_1 \,V_2^2 - 4\,V_1 ^3}{4\,\varDelta_6^{(4)}} 
= q^{-1} + 20 \,q - 62\, q^3 +216\, q^5 + \cO(q^6)\, .
\end{split}
\ee
It has a simple pole at the cusp $\infty$ and is regular at the cups $0$ and $1/2$,
\be
\begin{split}
 J_4 \big|\sigma_0 =& \frac{256}{J_4-8}+8 = 8 + 256 \, q + 2048\, q^2 + 11264\, q^3 + 49152\, q^4 + \cO(q^5)
 \,,\\
J_4 \big|\sigma_{1/2} =& -\frac{256}{J_4+8}+8 =8 - 256 \, q + 2048 \, q^2 - 11264\, q^3 + 49152\, q^4 +\cO(q^5)
\,.
\end{split}
\ee
The Hauptmodul of ${\rm SL}(2;\IZ)$ and $\Gamma_0(2)$ can be expressed in terms of $J_4$ as
\be
J=\frac{J_4^6-24 \,J_4^5+196992\, J_4^4+16770048\, J_4^3+377573376\,
   J_4^2+3220733952\, J_4+9396289536}{(J_4-8)^4\, (J_4+8)}\,,
\ee
and
\be
J_2 = J_4 + \frac{256}{J_4+8}\,.
\ee
Modular derivatives act as 
\be
\begin{split}
D \,V_1  = & -\tfrac13 \, V_1 \, \hat E_2 + 
\tfrac12\, V_2^2 - \tfrac13 \, V_1^2 - \tfrac12 \,V_1\, V_2 \,,
\\
D\, V_2 = & -\tfrac13 \, V_2 \, \hat E_2 -\tfrac12 \, V_2^2
+\tfrac53 \,V_1 \, V_2 \,,
\\
D\, [1/\varDelta_6^{(4)}] = & \frac{\hat E_2 - V_1 + \tfrac32 \,V_2}{ \varDelta_6^{(4)}}\,.
\end{split}
\ee

\subsection{$\Gamma_0(6)$  \label{appg06}}
The congruence subgroup $\Gamma_0(6)$ has index 12 in ${\rm SL}(2;\IZ)$ and 4 cusps at at $\infty$ , $0$ , $1/2$ and $1/3$. Its fundamental domain can be chosen as 
\be
\cF_6 = \{1,S,ST,ST^2,ST^3,ST^4,ST^5, ST^2 S, ST^2 ST, ST^2 ST^2, ST^3 S, ST^3ST
\}\,\cF\,.
\ee
The dimensions $d_w^{(6)}$ of $\cM_w (6)$ are given by the generating function
\be
\sum_{k\ge 0} d_{2k}^{(6)}\, x^{2k} = \frac{1+3x^2+4x^4+3x^6+x^8}{(1-x^4)(1-x^6)} = 1+3x^2+5x^4+7 x^6+9 x^8 +\dots\, . 
\label{dims06}
\ee
The ring of holomorphic modular forms under is generated by 
\be
U_1 = X_6(\tau)\,, \qquad U_2 = X_2(3\tau)\quad {\rm and} \qquad U_3 =X_3(2\tau)\,. 
\ee
In accordance with \eqref{dims06}, any element of $\cM_k(6)$ can be decomposed uniquely as 
\be
\begin{split}
\phi =&\, A + B_1 \, U_1 + B_2 \, U_2 + B_3 \, U_3 + C_1 \, U_1^2 + C_2 \, U_2^2 + C_3 \, U_3^2 + C_4 \, U_1\, U_2 
\\
&+ D_1\, U_1^3 +D_2\, U_2^3+D_3\, U_3^3+ E\, U_1^4 \,,
\end{split}
\ee 
where $A,B_i ,C_i ,D_i, E$ are holomorphic modular forms of ${\rm SL}(2;\IZ)$.
The first cusp form occurs at weight 4 and is given by
\be
\label{Del46}
\begin{split}
\varDelta_4^{(6)}  &=
\eta^2(\tau)\eta^2(2\tau)\eta^2(3\tau)\eta^2(6\tau)
\\
&=
\tfrac{1}{96} \left( -9 \, U_2^2 - 4 \, U_3^2  +6\, U_3  \, U_1 - U_1^2 +2 \, U_2 \, (U_1 -5 \, U_3 )\right)\\
&=  q - 2 \, q^2-3\, q^3+4\, q^4+6\, q^5 +6 \, q^6- 16\, q^7 + \cO(q^8)\,.
\end{split}
\ee
The scaling matrices $\sigma_0$, $\sigma_{1/2}$ and $\sigma_{1/3}$ associated to the cusps
 cusps 0, $1/2$ and $1/3$ act on the generators as 
\be
\begin{array}{llllll}
U_1 \big|\, \sigma_0 & = -U_1\,, &
U_1 \big|\, \sigma_{1/2} & = 3 \, U_2 - 2\,  U_3 \,, &
U_1 \big|\, \sigma_{1/3} & = 2 \, U_3  - 3 \, U_2 \,, 
\\
U_2 \big|\, \sigma_0 &= \tfrac13(2\, U_3-U_1)\,,&
U_2 \big|\, \sigma_{1/2} & = \tfrac{1}{3} \left(  U_1 - 2  \, U_3 \right) \,, &
U_2 \big|\, \sigma_{1/3} & = - U_2 \,, 
\\
U_3\big|\, \sigma_0 &= \tfrac12(3\, U_2 -U_1)\,,&
U_3 \big|\, \sigma_{1/2} & = - U_3 \,, &
U_3  \big|\, \sigma_{1/3} & = \tfrac{1}{2} \, \left( U_1 - 3 \, U_2  \right) \,,
\\
\hat E_2 \big|\,\sigma_0 &=\hat E_2 - U_1\,,&
\hat E_2 \big|\,\sigma_{1/2} &=\hat E_2 - U_3\,,&
\hat E_2 \big|\,\sigma_{1/3} &=\hat E_2 - U_2 \,.
\end{array}
\ee
In particular the weight-4 cusp form \eqref{Del46} is even under both $\sigma_{1/2}$ and $\sigma_{1/3}$.

The Hauptmodul for $\Gamma_0 (6)$ is the McKay-Thompson series associated to the conjugacy class $6E$ of the Monster,
\be
\begin{split}
J_6 = & \left(\frac{\eta(2\tau)\eta^3(3\tau)}{\eta(\tau) \eta^3(6\tau)}\right)^3 -3  
\\
 =& \frac{9 \, U_2^2 + 34\, U_2 \, U_3  - 4\, U_3^2 - 14\, U_2 \, U_1 + 16 \, U_3 \, U_1 - 3 \, U_1 ^2
 }{96\, \varDelta_4^{(6)}}
\\ 
=& q^{-1} + 6\, q+ 4\, q^2 - 3\, q^3 -12\, q^4 + \cO(q^5)
\end{split}
\ee
and has a simple pole at the cup at $\infty$. The behaviour near $0$, $1/2$ and $1/3$ is instead given by
\be
\begin{split}
J_6 \big| \sigma_0 &= \frac{72}{J_6-5}+5 = 5 + 72\, q + 360\, q^2 + 1368\, q^3 + 4392\, q^4 + \cO( q^5)
\,, 
\\
J_6 \big| \sigma_{1/2} &= \frac{9}{J_6+4}-4 = -4 + 9 \, q - 36\, q^2 + 90\, q^3 - 180 \, q^4 +\cO( q^5)
\,,
\\
J_6 \big| \sigma_{1/3} &= \frac{-8}{J_6+3}-3 = -3 - 8\, q + 24 \, q^2 - 24\, q^3 - 40 \, q^4 + \cO( q^5)
\,.
\end{split}
\ee
The Hauptmoduln of ${\rm SL}(2;\IZ)$, $\Gamma_0(2)$ and $\Gamma_0(3)$ can be expressed in terms of $J_6$ as
\be
\begin{split}
J_2=& J_6+\frac{270}{J_6+4}-\frac{972}{(J_6+4)^2}+\frac{729}{(J_6+4)^3}\,,
\\
J_3=&J_6+\frac{48}{J_6+3}+\frac{64}{(J_6+3)^2}\,,
\\
J = &  J_2 + J_3 - 2 J_6  + 432 \left[ 
\frac{455}{J_6-5}+ \frac{47484}{(J_6-5)^2} + \frac{1517184}{(J_6-5)^3}\right.
\\
&\quad \left.+\frac{21088512}{(J_6-5)^4}+ \frac{134369280}{(J_6-5)^5}+\frac{322486272}{(J_6-5)^6} \right]  \,.
\end{split}
\ee
Modular derivatives act as 
\be
\begin{split}
D \,U_1  =& -\tfrac13 \,U_1 \, \hat E_2 + \tfrac{1}{12}\,
 \left(-45\, U^2_2   -30 \, U_2\,   U_1+7 \,U_1^2\right)\,,
\\
D \,U_2 =& -\tfrac13\, U_2 \,\hat E_2
+\tfrac{1}{24} \,\left(9\, U^2_2  -2\, U_2 \,  (4\, U_3  +U_1)-4
   \,U^2_3  +U_1^2\right)\,,
\\
D \, U_3 =& -\tfrac13 \,U_3\,\hat E_2+
\tfrac{1}{48} \,\left(-99 \,U^2_2  -6 \,U_2  \, (4 \,U_3  +3 \,U_1)+5\, \left(4\,
   U^2_3  +U_1^2\right)\right)   \,,
   \\
D\, [1/\varDelta_4^{(6)}] = & \frac{ \tfrac23\, \hat E_2 +\tfrac{1}{6}\, (3 \,U_2\, +2 \, U_3 -3 \,U_1 ) }{ \varDelta_4^{(6)}}\,.
\end{split}
\ee

\providecommand{\href}[2]{#2}\begingroup\raggedright\endgroup
 
\end{document}